\documentstyle[12pt,psfig]{article}

\newcommand{\kms}{\mbox{km~s$^{-1}$}}
\newcommand{\mols}{\mbox{molec.~s$^{-1}$~}}

\renewcommand{\deg}{\mbox{$^{\circ}$}}
\newcommand{\arcsec}{\mbox{''}}

\begin{document}

\title{Radio observations of comet 9P/Tempel 1 before and after Deep Impact}

\author{Nicolas Biver$^a$
	Dominique Bockel\'ee-Morvan$^a$, J\'er\'emie Boissier$^a$, \\
	Jacques Crovisier$^a$, Pierre Colom$^a$,
	Alain Lecacheux$^a$, Rapha\"el Moreno$^a$,\\
{\small $^a$ LESIA, CNRS UMR 8109, Observatoire de Paris, 5 pl. Jules Janssen, F-92190 Meudon, France}\\
Gabriel Paubert$^b$,\\
{\small $^b$ IRAM, Avd. Divina Pastora, 7, 18012 Granada, Spain} \\
Dariusz C. Lis$^c$, Matthew Sumner$^c$,\\
{\small $^c$ California Institute of Technology, MC 320-47, Pasadena , CA 91125, USA}\\
Urban Frisk$^d$,\\
{\small $^d$ Swedish Space Corporation, PO Box 4207, SE-17104 Solna, Sweden}\\
{\AA}ke Hjalmarson$^e$, Michael Olberg$^e$, Anders Winnberg$^e$,\\
{\small $^e$ Onsala Space Observatory, SE-43992 Onsala, Sweden}\\
Hans-Gustav Flor\'en$^f$, Aage Sandqvist$^f$,\\
{\small $^f$ Stockholm Observatory, SCFAB-AlbaNova, SE-10691 Stockholm, Sweden}\\
and Sun Kwok$^{g,h}$\\
{\small $^g$ Dept. of Physics and Astronomy, University of Calgary, Calgary, AB T2N 1N4, Canada}\\
{\small $^h$ Inst. of Astron. \& Astrophys., Academia Sinica, PO Box 23-141, Taipei 106, Taiwan}}

\date{\today}
\maketitle
\noindent
Number of pages: 24 \\
Number of tables: 7 \\
Number of figures: 18 \\
\noindent
Proposed Running Head:\\
{Radio observations of the Deep Impact cometary target} \\
\\
Corresponding author: \\
Nicolas Biver\\
LESIA, Observatoire de Paris, \\
5 Place Jules Janssen \\
F-92190 Meudon \\
France \\
\\
E-mail: nicolas.biver@obspm.fr \\
Tel. 33 1 45 07 78 09 \\
Fax: 33 1 45 07 79 39 \\
\noindent
\newpage

\abstract{Comet 9P/Tempel 1 was the target of a multi-wavelength
worldwide investigation in 2005.  The NASA Deep Impact mission reached
the comet on 4.24 July 2005, delivering a 370~kg impactor which hit
the comet at 10.3~\kms.  Following this impact, a cloud of gas and
dust was excavated from the comet nucleus.  The comet was observed in
2005 prior to and after the impact, at 18-cm wavelength with the
Nan\c{c}ay radio telescope, in the millimetre range with the IRAM and
CSO radio telescopes, and at 557~GHz with the Odin satellite.

OH observations at Nan\c{c}ay provided a 4-month monitoring of the
outgassing of the comet from March to June, followed by the
observation of H$_2$O with Odin from June to August 2005.  The peak of
outgassing was found to be around $1\times10^{28}$ \mols between May
and July.  Observations conducted with the IRAM~30-m radio telescope
in May and July 2005 resulted in detections of HCN, CH$_3$OH and
H$_2$S with classical abundances relative to water (0.12, 2.7 and
0.5\%, respectively).  In addition, a variation of the HCN production
rate with a period of $1.73\pm0.10$ days was observed in May 2005,
consistent with the 1.7-day rotation period of the nucleus.  The phase
of these variations, as well as those of CN seen in July by Jehin et
al.  (2006), is consistent with a rotation period of the nucleus of
$1.715$ days and a strong variation of the outgassing activity by a
factor 3 from minimum to maximum.  This also implies that the impact
took place on the rising phase of the ``natural'' outgassing which
reached its maximum $\approx$4~h after the impact.

Post-impact observations at IRAM and CSO did not reveal a significant
change of the outgassing rates and relative abundances, with the
exception of CH$_3$OH which may have been more abundant by up to one
order of magnitude in the ejecta.  Most other variations are linked to
the intrinsic variability of the comet.  The Odin satellite monitored
nearly continuously the H$_2$O line at 557~GHz during the 38 hours
following the impact on the 4th of July, in addition to weekly
monitoring.  Once the periodic variations related to the nucleus
rotation are removed, a small increase of outgassing related to the
impact is present, which corresponds to the release of
$\approx5000\pm2000$ tons of water.  Two other bursts of activity,
also observed at other wavelengths, were seen on 23 June and 7 July;
they correspond to even larger releases of gas.  } \\



\section{Introduction}


The investigation of the composition of cometary nuclei is crucial for
understanding their origin.  Short-period Jupiter-family comets may
have accreted directly in the Kuiper Belt beyond Neptune.  Having
spent most of their time in a very cold environment, these objects may
not have evolved very much since their formation.  Thus, their
composition might provide direct clues to the composition in the outer
regions of the Solar Nebula where they formed.

However, after spending many decades in a warmer environment within 6
AU from the Sun, the upper layers of such cometary nuclei may have
evolved, probably loosing their most volatile components, and one may
wonder if material released in the coma at the time of perihelion is
representative of the bulk composition of the comet.  One key
objective of the Deep Impact mission (A'Hearn et al.\ 2005a) was to
excavate material from a depth of a few tens of metres in order to see
the release of probably more pristine material.  The target, comet
9P/Tempel~1, is on 5.5-year orbit and belongs to the Jupiter-family
group of comets, like 19P/Borrelly that was investigated in 2001
(Bockel\'ee-Morvan et al.\ 2004) when Deep Space 1 passed within 2300
km of its nucleus.  
The Deep Impact spacecraft released a 370~kg impactor which hit
the comet nucleus at 10.3~\kms on 4.2445 July 2005 UT Earth-based time
(A'Hearn et al.\ 2005b).  Following this technological success, a large
cloud of dust and icy particles flew out of the impact crater (at an
average speed of 0.2~\kms; Keller et al.\ 2005) which was the target
of spectroscopic investigations.  The objective was to characterize the
amount and composition of material excavated and to compare it with
the chemical composition of the comet before the impact.

Observing comet 9P/Tempel 1 in support to the Deep Impact mission
(Meech et al.\ 2005a) was a major objective of ground based radio
observations and a major objective for the 2005 Odin space observatory
program (Nordh et al. 2003).  We report in the present paper our 
observations conducted with the Nan\c{c}ay radio telescope, the IRAM 
(Institut de radioastronomie millim\'etrique) 30-m telescope, the CSO 
(Caltech Submillimeter Observatory), and with the Odin
satellite.


\section{Observations}
	Comet 9P/Tempel 1 returned to the perihelion of its orbit 
on 5 July 2005, just one
day after it was hit by the projectile released by the Deep Impact spacecraft. 
Our radio observing campaign started 4 months earlier, in March--May, 
when the comet was more favorably placed in the sky. Opposition was indeed 
on 4 April and perigee was on 4 May 2005 at 0.712 AU from the Earth. 
Comet 9P/Tempel 1 was also known to have a peak
in activity about two months before perihelion (Lisse et al. 2005).
The first IRAM observing campaign to characterize the comet activity and
chemical composition was scheduled in May 2005. 
The OH maser inversion was maximum ($i<-0.20$) in March to May
2005, which made it the best observing window for observations at Nan\c{c}ay.
On the other hand, solar elongation constraints (60\deg$<elong.<120$\deg) 
prevented Odin observations before 7 June 2005.
The second observing campaign involving IRAM~30-m,
CSO, Nan\c{c}ay and Odin observatories was scheduled around 
the time of the impact (early July 2005).

\subsection{Nan\c{c}ay radio telescope}

\begin{figure*}\vspace{-0.5cm}\hspace{1cm}
   \psfig{width=15cm,angle=0,figure=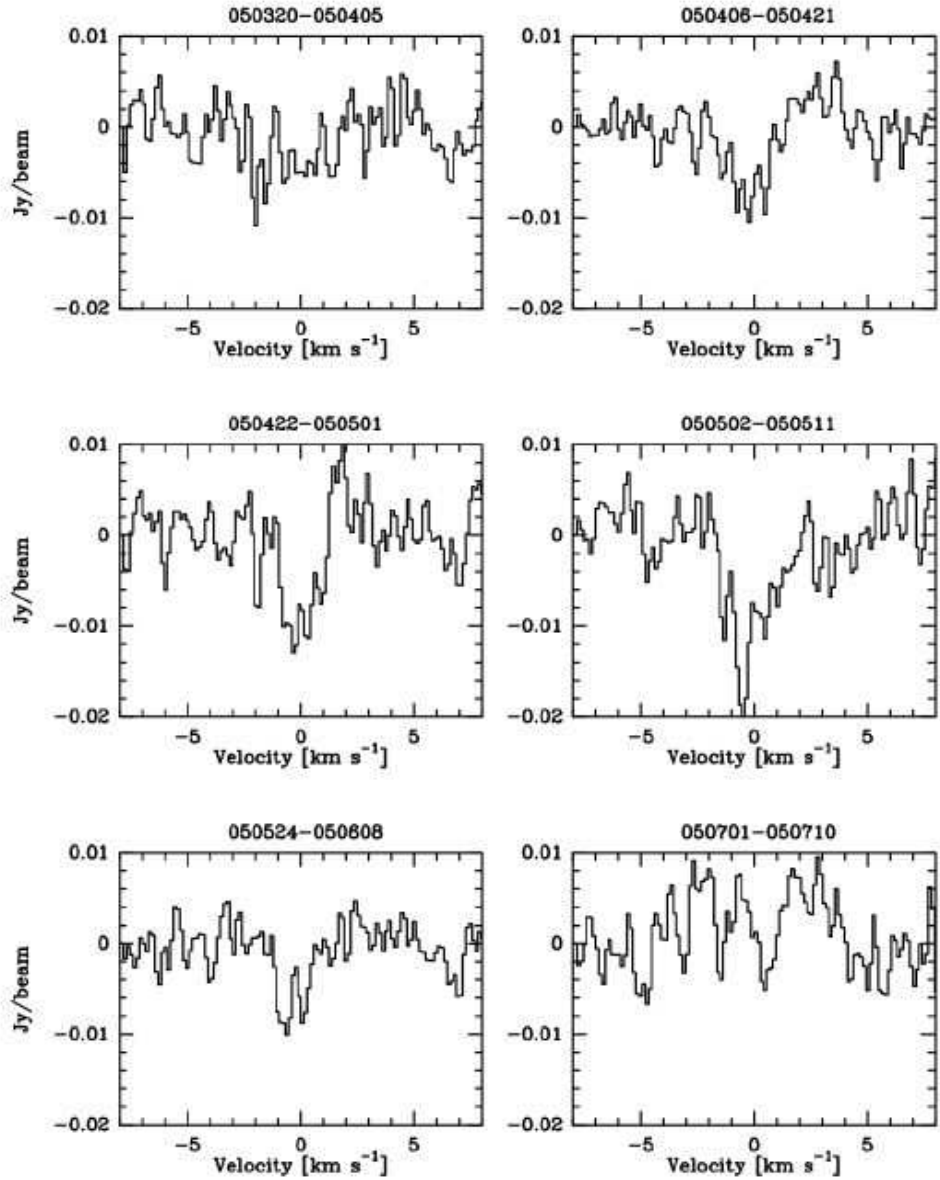}\vspace{-0.5cm}
   \caption{Selected spectra of the OH radical observed in
	comet 9P/Tempel~1 with the Nan\c{c}ay radio telescope.}
   \label{fig9poh}
\end{figure*}

The Nan\c{c}ay radio telescope is a meridian telescope with a
fixed primary spherical mirror (35 $\times$ 300~m), a secondary plane
mirror (40 $\times$ 200~m) tiltable on an horizontal axis and a focal
system that can track the source during approximately one hour around
time of transit.  For the observations of comet 9P/Tempel~1,
transiting at $51\deg$ to $32\deg$ elevation, the beam size at 18-cm 
wavelength is 3.5$\times$19'.  Both polarizations of the two hyperfine
transitions of the OH $\Lambda$-doublet at 1665 and 1667 MHz were
observed every day in the comet from 4 March to 8 June and from 1 to
10 July 2005.  The comet was too weak to be detected on a single day
of observation but was detected on averages of about 10 days since 20
March (Crovisier et al.\ 2005).  The observations are listed in
Table~\ref{tabobsoh} and examples of the observed spectra are shown in
Fig.~\ref{fig9poh}.

Observations were reduced and production rates calculated as has been
regularly done for previous observations (Crovisier et al.\ 2002),
assuming isotropic outgassing and taking into account collisional
quenching of the maser (Table 4).  The UV pumping by the solar
radiation field is responsible for the population inversion $i$ of the
ground state $\Lambda$-doublet of OH that makes the comet emission
detectable (Despois et al.\ 1981; Schleicher \& A'Hearn 1988).  The
Swings effect makes $i$ very sensitive to the heliocentric velocity of
the comet.  In July $i$ was close to zero (+0.03 or +0.08 on average,
depending upon the inversion model).  This partly explains why the
comet was not detected at Nan\c{c}ay at that time.  The Nan\c{c}ay
upper limit is consistent with the water production rates observed by
other means (see below) and with the OH signal detected with the Green
Bank Telescope (Howell et al. 2005).

\bigskip

[Table~\ref{tabobsoh}]

\bigskip

[Fig.~\ref{fig9poh}]


\subsection{IRAM~30-m telescope}
	Comet 9P/Tempel 1 was first observed between 4.8 and 9.0 May 2005 with
the IRAM~30-m radio telescope, around the time of its perigee. The weather 
was relatively good and stable, with higher sky opacity 
on the first night. On the last scheduled night (9.8--10.0 May) observations 
had to be stopped due to strong winds. 
In July, the comet was observed every evening from 2.7 to 10.8
July 2005: 6 to 8 hours of observations were scheduled per day, but in 
general the stability of the atmosphere was relatively poor during the first 
3 hours, leading to a typical pointing uncertainty of 5--6\arcsec~.
Later in the evening pointing stability and atmospheric transmission usually
improved. The best observing conditions were on 2 and 4 July evenings, while
the weather on 6, 8, 9 and 10 July was relatively poor (7 to 12mm of
precipitable water versus 2 to 5mm).
Every night in May and July, Jupiter and the carbon star IRC+10216 were 
observed for calibration purpose. The beam sizes and main beam efficiencies
($\eta_b\approx0.76$ to 0.40 depending on frequency) were checked on  
planets (Jupiter, Saturn or Venus).
The calibration of the line intensity and pointing 
stability were evaluated from observation of the compact source IRC+10216.
In addition, other pointing sources within 10--15\deg~ of the comet were 
observed at least every hour.

The observations of the HCN~$J$(3--2) line in IRC+10216 were especially useful
to estimate the effect of atmospheric instability on the comet line intensity.
Anomalous refraction (Altenhoff et al. 1987), which causes  
random pointing offsets, was the main cause of signal losses in July.
Indeed, at this frequency the antenna half power beam width is 9.4\arcsec
which makes observations very sensitive to pointing. 
In May, the mean standard deviation of the daily measurements of the intensity 
of the HCN~$J$(3--2) line in both IRC+10216 and the Orion Molecular Cloud 
calibration sources was less than 7\%.
But a drop of 50\% of the HCN~$J$(3--2) line intensity 
of IRC+10216 was often observed in July at the beginning of the 
observations, while the losses decreased to less than 20\% at the
end of each evening shift. The effect on the HCN~$J$(1--0) line intensity
was much smaller (less than 10\% variation) due to the larger beam. 
These variations could be well attributed to an average pointing offset
-- as provided in Table~\ref{tabobs} -- which was larger at the beginning 
of the observations.

\begin{figure*}\vspace{-0cm}\hspace{1cm}
   \psfig{width=15cm,angle=0,figure=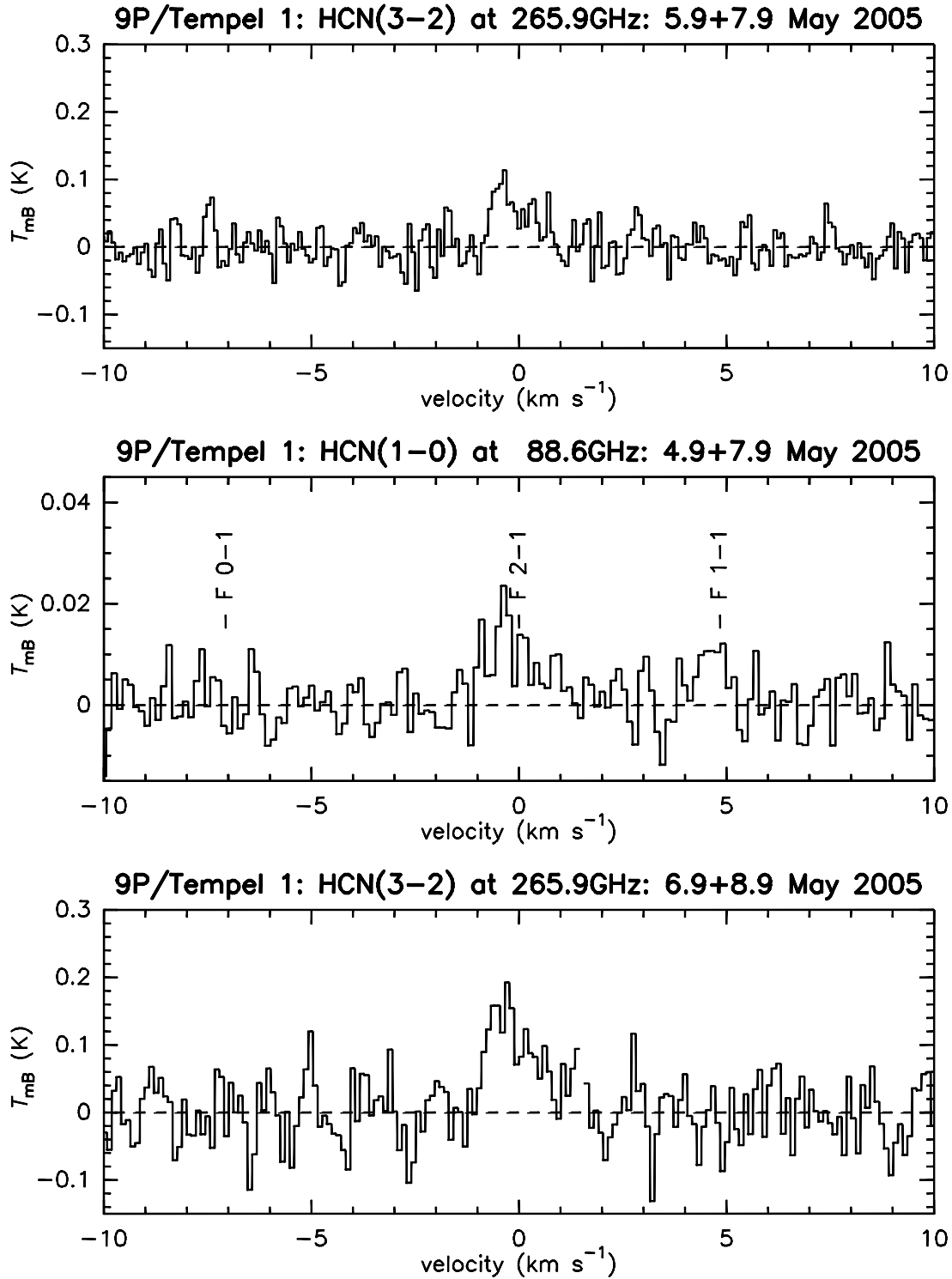}\vspace{-0.5cm}
   \caption{Spectra of HCN obtained with IRAM in May 2005.
	These spectra are two-day averages corresponding to the minimum
	outgassing rate (HCN~$J$(3--2), top), 
	an intermediate case (HCN~$J$(1--0), middle)
	and the maximum outgassing rate (HCN~$J$(3--2), bottom).}
   \label{fig9phcnjet}
\end{figure*}

\begin{figure*}\hspace{1cm}
   \psfig{width=15cm,angle=270,figure=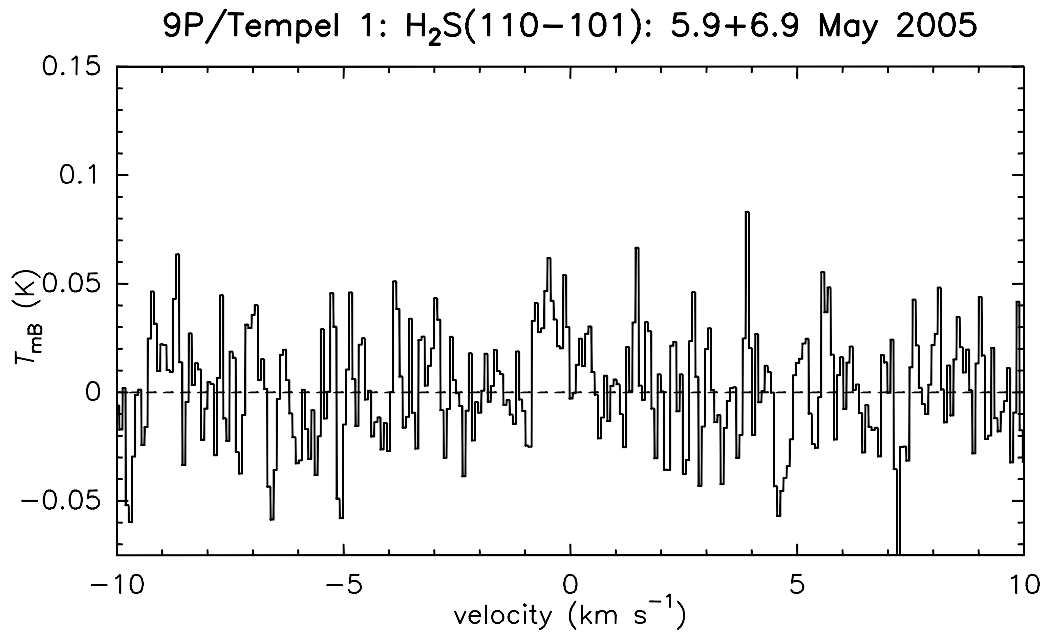}\vspace{-0.5cm}
   \caption{IRAM~30-m telescope spectrum of H$_2$S $1_{10}-1_{01}$ line at 168.8~GHz
	observed in May 2005.}
   \label{figh2s}
\end{figure*}

\begin{figure*}\hspace{1cm}
   \psfig{width=15cm,angle=270,figure=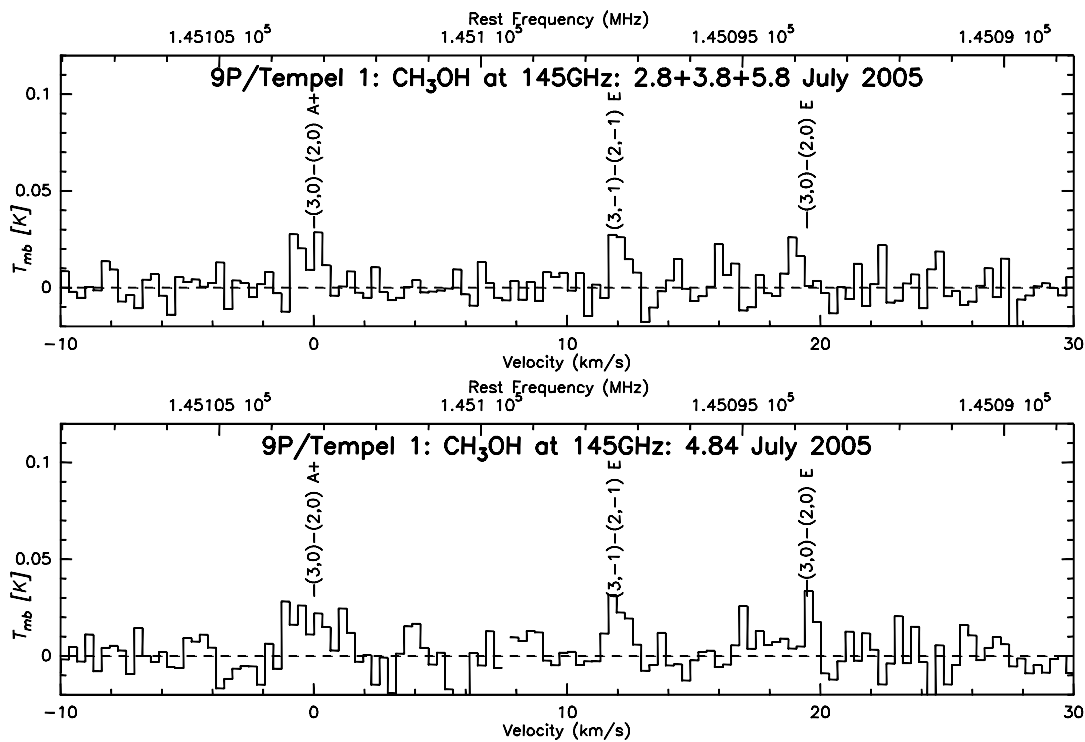}\vspace{-0.5cm}
   \caption{IRAM~30-m telescope spectra of methanol lines at 145.1~GHz
	on the impact day (4.84 July, bottom) and the average of 
	other days not affected by the impact ejecta (top).}
   \label{figch3oh}
\end{figure*}

\begin{figure*}\hspace{2cm}
   \psfig{width=12cm,angle=0,figure=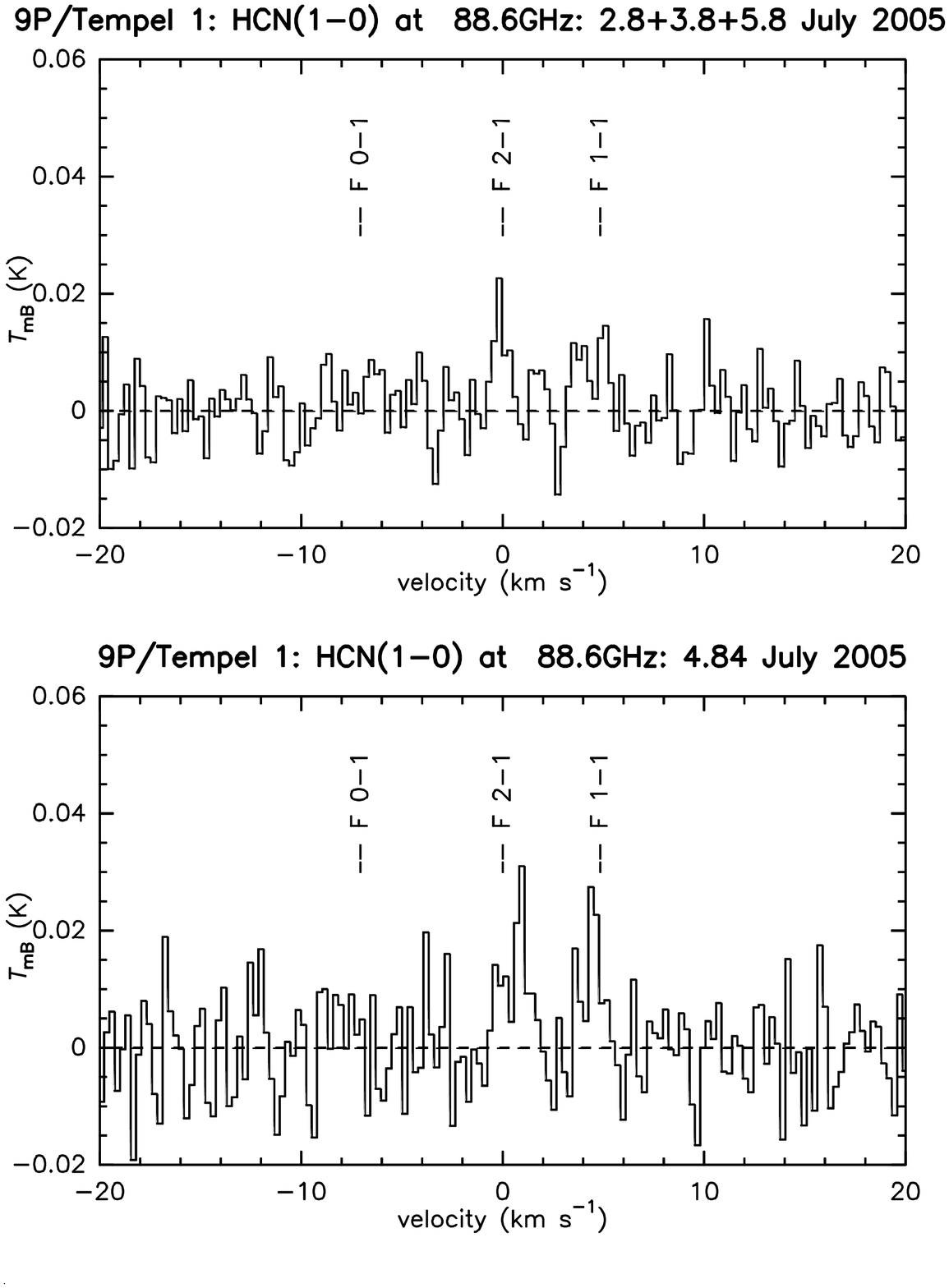}\vspace{-0.5cm}
   \caption{IRAM~30-m telescope spectra of the HCN~$J$(1--0) line on
	the impact day (4.84 July, bottom), and the average of 
	other days not affected by the impact ejecta (top).}
   \label{fighcn1}
\end{figure*}

\begin{figure*}\hspace{2cm}
   \psfig{width=12cm,angle=0,figure=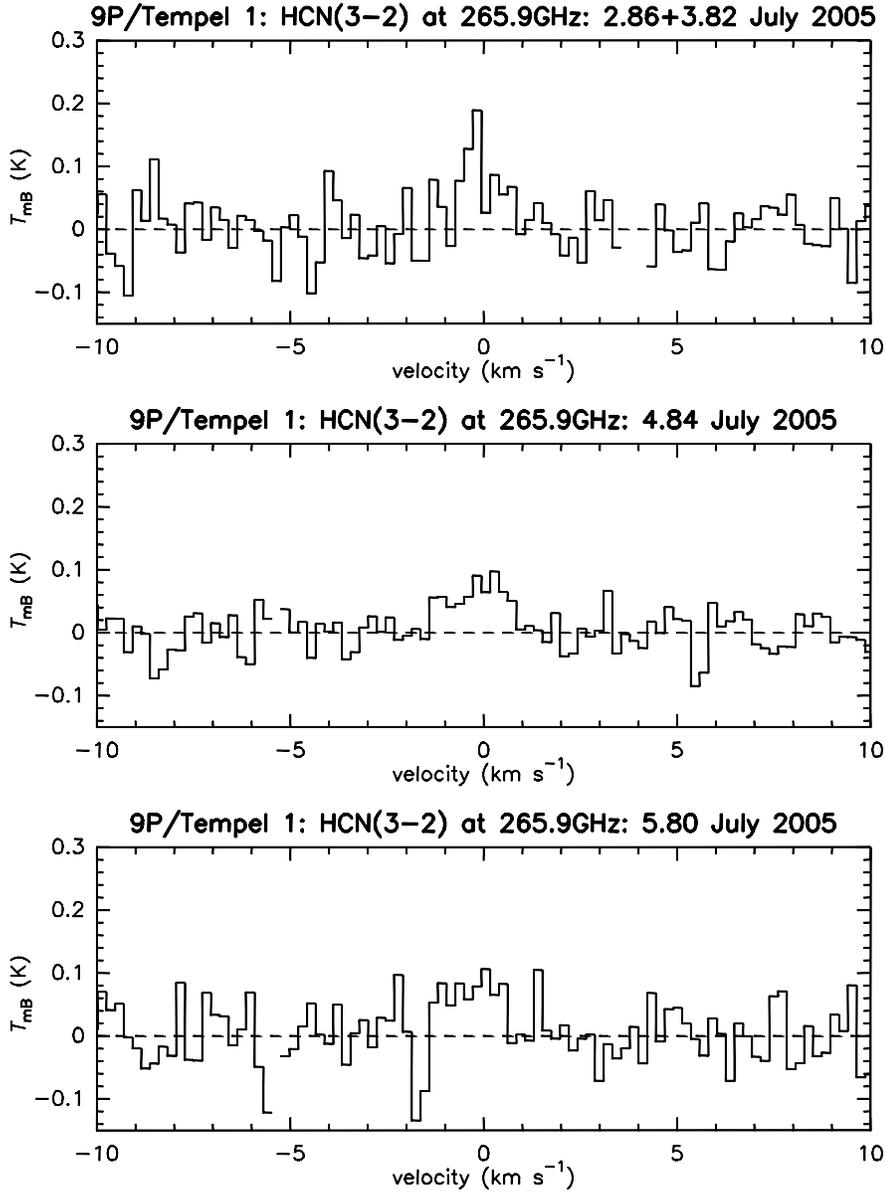}\vspace{-0.5cm}
   \caption{IRAM~30-m telescope spectra of HCN~$J$(3--2) line
	on the impact day (4.84 July, middle), on the day after (bottom) 
	and average of the two previous days not affected by the impact ejecta (top).}
   \label{fighcn3}
\end{figure*}

HCN~$J$(1--0) and HCN~$J$(3--2) lines were detected both in May and July. 
Average spectra are shown in Figs.~\ref{fig9phcnjet},\ref{fighcn1} 
and~\ref{fighcn3}.
Evidence of time variability in the HCN line intensities that could be 
attributed to the rotation of the nucleus was readily seen in May 
(Biver et al. 2005). H$_2$S (Fig.~\ref{figh2s}) and CH$_3$OH were also 
marginally detected in May. Up to three individual lines of CH$_3$OH lines at 
145~GHz were detected in July (Fig.~\ref{figch3oh}). Other species 
(CS, CO, H$_2$CO) were searched for both in May and July but not detected.
Line intensities or 3-$\sigma$ upper limits are reported in Table~\ref{tabobs}.

\bigskip
[Fig.~\ref{fig9phcnjet}]
\bigskip
[Fig.~\ref{figh2s}]
\bigskip

\bigskip
[Figs.~\ref{figch3oh}]
\bigskip
[Figs.~\ref{fighcn1} and ~\ref{fighcn3}]
\bigskip

\subsection{CSO telescope}
In contrary to IRAM~30-m, the CSO with 10.4-m telescope on top 
Mauna-Kea, Hawaii, was in direct viewing of comet 9P/Tempel 1 at the
time of the encounter with Deep Impact. Although benefiting from a better
sky transparency (average amount of precipitable water was 1--2~mm versus
3--4~mm at IRAM on 4--5 July), this 10.4-m radio telescope has a lower
sensitivity than the IRAM~30-m and could only observe during 4~h 
between sunset and comet set. Observations were targeted on the day of 
the impact and the next one.
The 345~GHz double-sideband receiver was used and tuned to the pair of 
methanol lines at 304.2 and 307.2~GHz, predicted to be among the strongest 
lines in this wavelength domain (e.g. Biver et al. 2000). In addition, 
they can provide precise information on the gas temperature. 
Observations started at 5~h UT on 4 July until 9~h UT
and from 1.5 to 8~h UT on 5 July 2005 with tuning to the HCN~$J$(4--3) line 
at 354.5 GHz after 6 UT. Jupiter was used as a pointing and 
beam efficiency calibration source ($\eta_b\approx0.6$ at 305~GHz and 353~GHz).

\begin{figure*}\hspace{1cm}
   \psfig{width=15cm,angle=270,figure=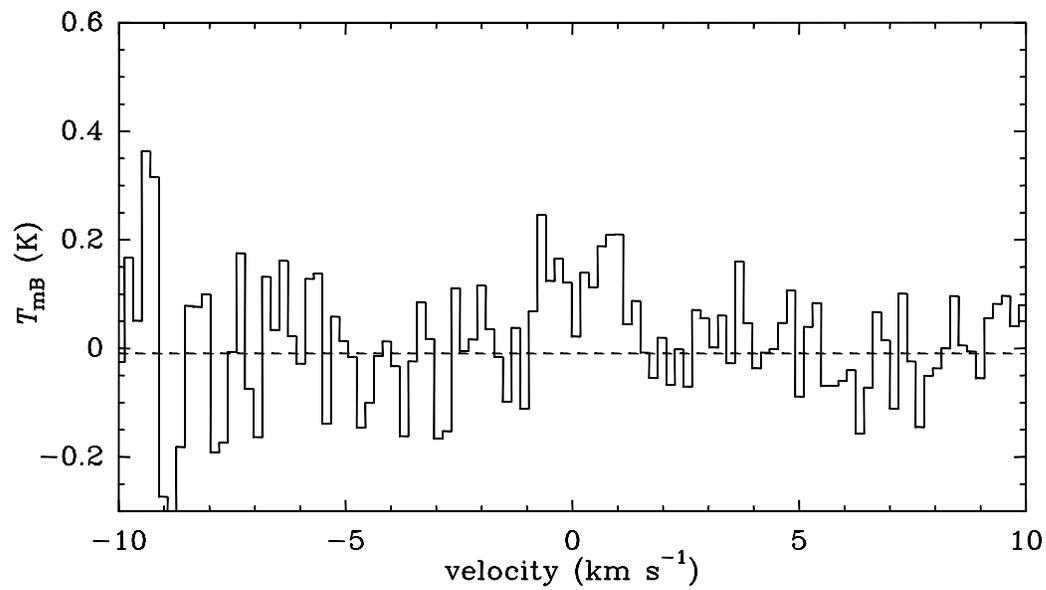}\vspace{-0.5cm}
   \caption{CSO spectrum of the sum of the two methanol 
	lines at 304.2 and 307.2~GHz. This spectrum is an average
	of 2.5~h of observations following the impact on 4.24 July 2005.}
   \label{figch3ohcso}
\end{figure*}

A marginal detection of the CH$_3$OH lines at 304/307~GHz 
(4-$\sigma$, Fig~\ref{figch3ohcso})
was obtained on the 4.3 July post-impact data after the two lines
(expected to have similar intensities) were co-added. HCN was not detected. 
Line intensity and upper limits are given in Table~\ref{tabobs}. 

\bigskip
[Figs.~\ref{figch3ohcso}]
\bigskip

\subsection{Odin satellite}

The Odin satellite (Nordh et al. 2003, Frisk et al. 2003) 
houses a 1.1-m telescope
equipped with 5 receivers at 119~GHz and covering the 486--504~GHz and 
541--580~GHz domains that are in large part unobservable from the ground.
Half of the time is dedicated to astronomical studies and the other
half to aeronomical investigations. Comets are a major topic for Odin 
observations and the fundamental rotational line of water at 556.936~GHz
has been detected in 11 comets between 2001 and 2005. In addition 
H$_2^{18}$O at 547.476~GHz and NH$_3$ at 572.498~GHz were detected
in some comets (Biver et al. 2006b).

Comet 9P/Tempel 1 was observed during the ``eclipse period'', when part of 
Odin orbit lies in the Earth shadow, so that only one receiver at 557~GHz 
could be used due to power limitations. The single-side band receiver
system temperature ($Tsys$), which is measured three times per orbit, was 
$3100\pm30$ K in June, $3200\pm60$ K in early July 
and about 3350~K at the end of July and beginning of August.  
Variations of $Tsys$ with time were smooth and the resulting calibration 
uncertainty was below 1\%. Two spectrometers were used: the
wide band Acousto-Optical Spectrometer (AOS) (1~GHz with 0.6~MHz 
channel spacing and 1~MHz resolution) and the high resolution 
AC2 autocorrelator (112~MHz bandwidth, 125~kHz channels,
333~kHz resolution after Hanning smoothing; Olberg et al. 2003). The high 
resolution corresponds to 180~m~s$^{-1}$ at 557~GHz, well enough to resolve 
the $\approx$1.5~\kms~ wide cometary line. 95 orbital revolutions of 1.6~h,
in series of 4--5 consecutive ``orbits'', were 
allocated to the 9P/Tempel 1 program. Due to orbital constraints the comet was 
observed only 45 to 55~min per revolution. The observations during the 
ten ``orbits'' scheduled on 7/8 June mostly failed due to an error in the  
setup of the spectrometers while three ``orbits'' on the 17 June and 
two on the 3rd of July were pointed too far from the target. 
A 2$\times$2' map was obtained on 18 June, while on other dates the 
observations were aimed at the comet nucleus position. The residual 
pointing offset (checked on Jupiter maps)
and uncertainty are estimated to be less than 20''. The beam size at 
557~GHz is 127'' and main-beam efficiency $\eta_b=0.85$. 
Due to the weakness of the comet, the signal was 
averaged over 4 to 6 ``orbits''. This offers a time resolution (6--10~h)
and sensitivity to the amount of gas released by the impact which are better 
than what the submillimeter wave satellite (SWAS) could achieve 
(Bensch et al. 2006).
Due to standing waves and other interferences, several
ripples were present in the spectra and removed with sinusoidal baseline
fitting. The line intensities and velocity shifts given in 
Table~\ref{tabobs} are the average of AC2 and AOS values.

\bigskip
[Table~\ref{tabobs}]
\bigskip

\section{Data analysis}

\subsection{Molecular production rates}
	Intensities of millimetre and submillimetre lines were converted 
into production rates using the same modelling and codes  
as in previous papers (e.g. Biver et al. 1999, 2000, 2002, 2006a, 2006b). 
A Haser model with symmetric outgassing -- unless specified (Section 4.2) 
-- and constant radial expansion velocity are used to describe the density, 
as in our previous studies. The variation 
of the photo-dissociation lifetimes of H$_2$O and HCN due to solar
activity is taken into account. 

	The expansion velocity ($v_{exp}$) is 
estimated from the line shapes of the spectra with highest 
signal-to-noise ratios. For May observations we use $v_{exp}=0.65$~\kms~ 
(half widths at half maximum intensity, $HWHM = 0.70\pm0.05$ 
and $0.67\pm0.13$~\kms~for the HCN~$J$(3--2) blue and red-shifted sides). 
In June--July there is more scatter in the measurements: 
$HWHM = 0.78$ and $0.69\pm0.05$~\kms~ for H$_2$O,
$HWHM = 1.00$ and $0.58\pm0.15$~\kms~ for HCN~$J$(3--2) and 
$HWHM = 0.75$ and $0.41\pm0.11$~\kms~ for HCN~$J$(1--0), 
for blue and red-shifted sides of the lines, respectively, and we adopted an
average value of $v_{exp}=0.75$~\kms.
We used 0.65~\kms~ for H$_2$O data obtained after mid-July when lines were 
narrower
($HWHM = 0.77$ and $0.42\pm0.07$~\kms~ for H$_2$O).
The variation of the line shape and expansion velocity will be discussed
in Sections 4.2 and 4.3.

Our model of excitation of the rotational levels of the molecules 
takes into account
collisions with neutrals at a constant gas temperature.
Collisions with electrons also play a major role and are modelled
according to Biver (1997) and Biver et al. (1999) with an electron 
density multiplying factor $x_{ne}$ set to 0.3 for all data. 
The electron density obtained with a factor $x_{ne}=0.2$--0.3 was
found to provide the best match to the radial evolution of line intensities
observed for several comets in extended Odin H$_2$O maps (Biver et al. 2006b). 
Interpretation of millimetre data (Biver et al. 2006a) usually requires a 
slightly higher factor, around 0.5. Data obtained in 9P/Tempel 1 cannot
put stringent constraints on this free parameter, but the coarse map
of H$_2$O($1_{10}-1_{01}$) obtained in June is as well matched with intensities computed
with $x_{ne}$ from 0.1 to 1.0. We adopted the mean value of $x_{ne}=0.3$: 
using $x_{ne}=0.2$ would increase systematically all H$_2$O production rates by 7\% and 
using $x_{ne}=0.5$ would decrease them by 10\%. 
The optical thickness of the water rotational lines is taken into account into 
the excitation process using the Sobolev ``escape probability'' 
method (Bockel\'ee-Morvan 1987).
	We have weak constraints on the gas temperature: in May, from 
simultaneous observations of the two HCN~$J$(3--2) and 
$J$(1--0) lines, we infer a rotational temperature of $12.2\pm1.1$ K.
Given the assumed collision rate, this corresponds to kinetic 
temperature of $14^{+11}_{-8}$K. For July data we do not expect
a much higher temperature (given that the generally observed trend in previous
comets is $T\propto r_h^{-1}$ to $r_h^{-1.5}$) and assume $T = 20$ K.
This is slightly lower than the kinetic temperature estimated 
from infrared spectroscopy by Mumma et al. (2005) but this discrepancy
has always been observed when comparing radio and infrared measurements, 
which are made with different instrumental fields of view.
A temperature of only 18~K was measured in the more active 
comet C/1996 B2 (Hyakutake) at $r_h=1.55$ AU (Biver et al. 1999).   

	A radiative transfer code, taking into account line optical thickness,
is used to compute line intensities and to simulate line profiles. The mean 
opacity of the 557~GHz water line is $<$ 1 and its line intensity is almost 
proportional to the water production rate for values in the range 
5--12$\times10^{27}$~\mols. Other lines considered in this paper are 
optically thin.

\subsection{Simulations of time-variable outgassing rates}

	A caveat of radio observations is their large 
field of view, which samples molecules 
released by the nucleus at different times. The smaller the beam, 
the more likely the signal will reflect the outgassing behavior 
of the nucleus in real time. Since we cannot retrieve time dependent
information on the true nucleus outgassing rate directly or in a 
simple way, the best is to try to simulate the observations with a
minimum number of free parameters. For all observations and simulations, unless
specified, line intensities are converted into ``apparent'' production
rates, i.e., the production rates which would give the observed line 
intensities assuming a stationary regime. Apparent production rates
are computed following the description given in Sect. 3.1. 

\subsubsection{Simulation of periodic fluctuations}

First, we investigate the effect of a periodic fluctuation in outgassing rate 
in order to interpret the periodicities observed in HCN line intensities 
in May and, more marginally, in H$_2$O measurements obtained in July (Sect. 4),
both related to nucleus rotation.  
We assumed an outgassing rate with a sinusoidal periodicity 
of 1.7 days and an amplitude of 50\% of the mean value.
This periodic variation is accounted for when computing collisional 
excitation and radiation transfer. Isotropic outgassing, constant 
velocity and temperature are otherwise assumed. The expected line 
intensities for May and July observing conditions were computed
for various phases. They were then converted into apparent production
rates for direct comparison to the input true production rate.
The result is a quasi sinusoidal variation of the apparent production rate
which amplitude and phase are retrieved from the fit of a sinusoid.
This provides us with information on the extent to which the amplitude of
variation is reduced due to beam averaging, and the periodic variation is
shifted in phase (the phase shift is hereafter called ``beam delay''). 
As test cases, we studied the HCN~$J$(1--0) 
and HCN~$J$(3--2) lines observed at IRAM (which encompass the whole range of 
CSO and IRAM beam sizes (9.5--27\arcsec)) and H$_2$O observed with Odin.

\begin{figure*}\vspace{-0.cm}\hspace{1cm}
    \psfig{width=15cm,angle=270,figure=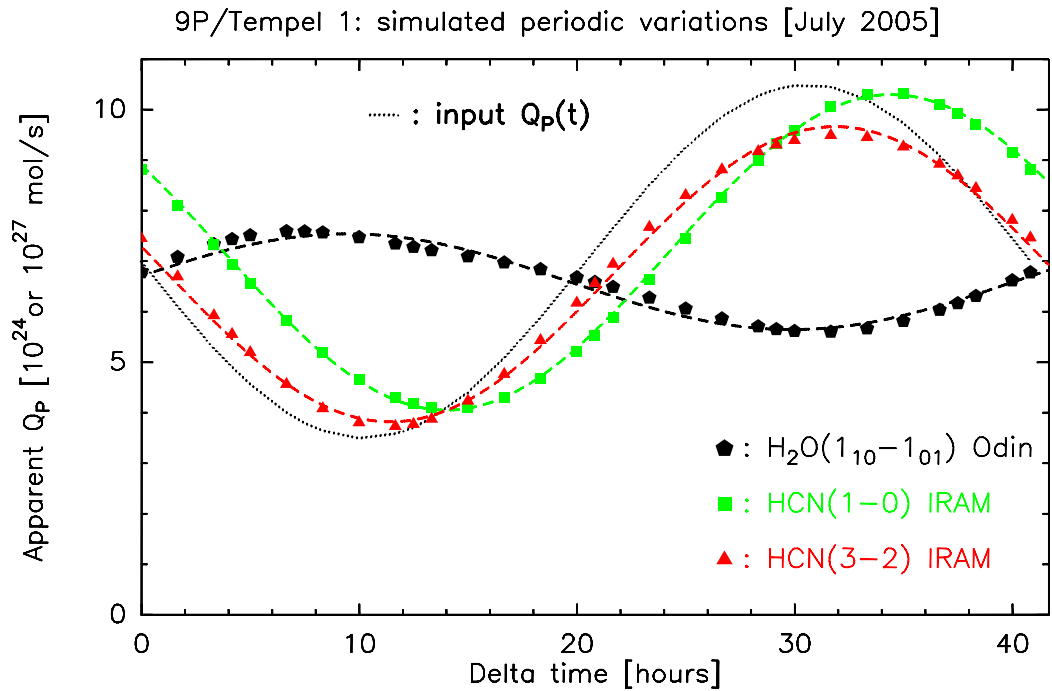}\vspace{-0.5cm}
   \caption{Simulation of the effect of a sinusoidal variation of the
	HCN and H$_2$O production rates (dotted line) on apparent
	production rates measured on different lines with Odin or IRAM 
        for the July observing conditions. 
	Dashed lines are the least squares sinusoidal fits to the computed values.
	This plot shows the phase shifts (``beam delay'') and 
	amplitude attenuations due to the corresponding beam sizes
	of the instruments. 
	The simulation for HCN observations in May yields very similar results.}
   \label{figsimrot}
   \end{figure*}

Figure~\ref{figsimrot} shows the model line intensities converted into
``apparent'' production rates. The periodicity expected in signals 
observed at IRAM is delayed by 1.1~h (HCN~$J$(3--2)) to 3.6~h (HCN~$J$(1--0)), 
and the delay for H$_2$O observed by Odin in July is 19.7~h.
The amplitude is reduced by 14\% (HCN) to 70\% (H$_2$O). 
Since observations are averaged over a few hours, the amplitude of variation
is further reduced by a
factor $\sin(x)/x$ with $x=2\pi t_o/T_p$, where  $T_p=1.7$ days and 
$t_o$ is the observation
duration (typically 0.2~day for HCN to 0.36~day for H$_2$O; 
Table~\ref{tabobs}). This effect is small for IRAM 
observations (amplitude typically reduced by 9\%) but significant
for Odin observations (27\%). Overall, the amplitude of variation of 
apparent H$_2$O production rates measured by Odin can be expected to be 
3.5 times smaller than the amplitude of variation of HCN apparent production 
rates, if H$_2$O and HCN productions 
presented similar periodic fluctuations at the nucleus surface.

For OH at Nan\c{c}ay, intensity variations related to nucleus rotation 
cannot be detected given the large beam size and the low signal-to-noise 
ratios obtained daily.

\bigskip
[Fig.~\ref{figsimrot}]
\bigskip

\subsubsection{Simulations of outgassing bursts}

We investigate here the lines intensity response to an outburst.
This study is focussed on the lines actually observed with radio facilities
in July 2005 and ``H$_2$O($1_{10}-1_{01}$)'' refers to this line seen 
with Odin beam and ``HCN~$J$(3--2)'' and ``HCN~$J$(1--0)'' refer hereafter to 
those lines as observed with IRAM beams (9.4\arcsec and 26.9\arcsec 
respectively).
The outburst is described by a  
sudden increase $\Delta Q$ of outgassing which then decreases 
following a half Gaussian with a characteristic time of half decay $\Delta t$.
For the test cases, we assumed a total release of material during the outburst 
$N_t$ on the order of $1.3\times10^{32}$ molecules.
The background production rate of water used is 
$Q_{m\rm H_2O}=8\times10^{27}$~\mols.

\begin{figure*}\vspace{-1cm}\hspace{1cm}
   \psfig{width=15cm,angle=270,figure=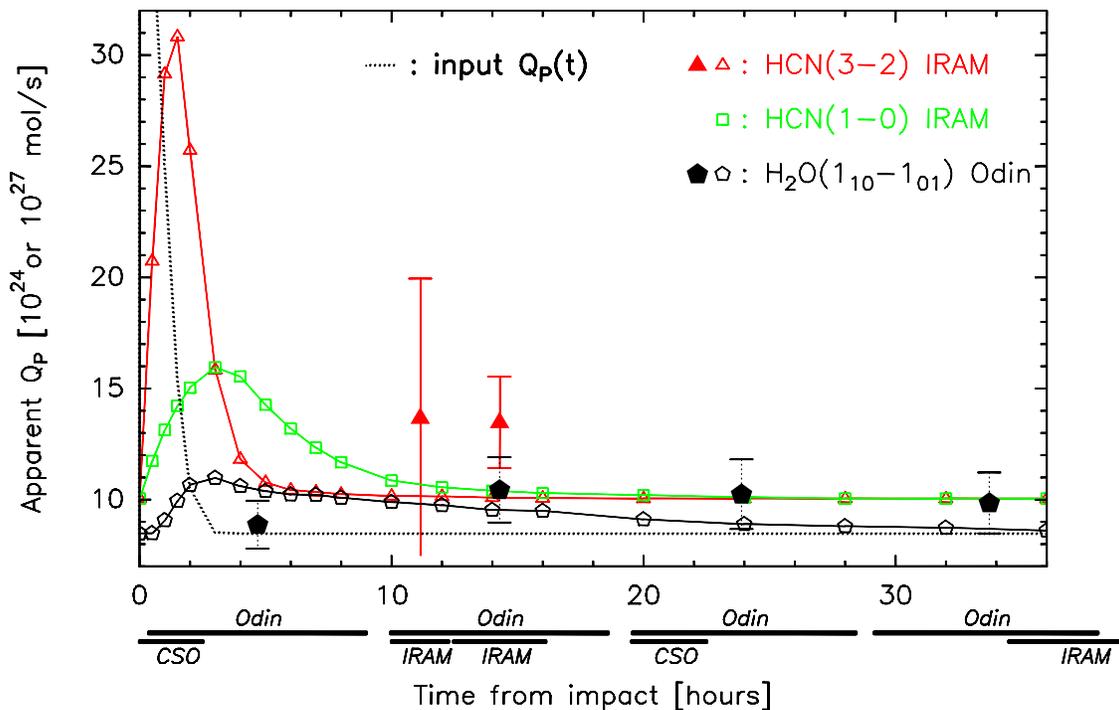}\vspace{-0.5cm}
   \caption{Simulation of the effect of an outburst of $1.3\times10^{32}$ water
	molecules ($0.12\times1.3\times10^{30}$ for HCN) released isotropically 
        in 1~h, at a velocity of 0.75~\kms (dotted line). 
	Connected open symbols are the apparent simulated  
	production rates, as in Fig.~\ref{figsimrot}.
	The observing time intervals of Odin, IRAM and CSO after 4.244 July
	are depicted below the bottom axis with horizontal bars.
	Measurements ($8\times10^{27}$~\mols + excess of outgassing given in
	Table~\ref{tabqh2oimp} for water) 
	are plotted with filled symbols and error-bars.}
   \label{figsimb1h}
\end{figure*}

\begin{figure*}\vspace{-1cm}\hspace{1cm}
   \psfig{width=15cm,angle=270,figure=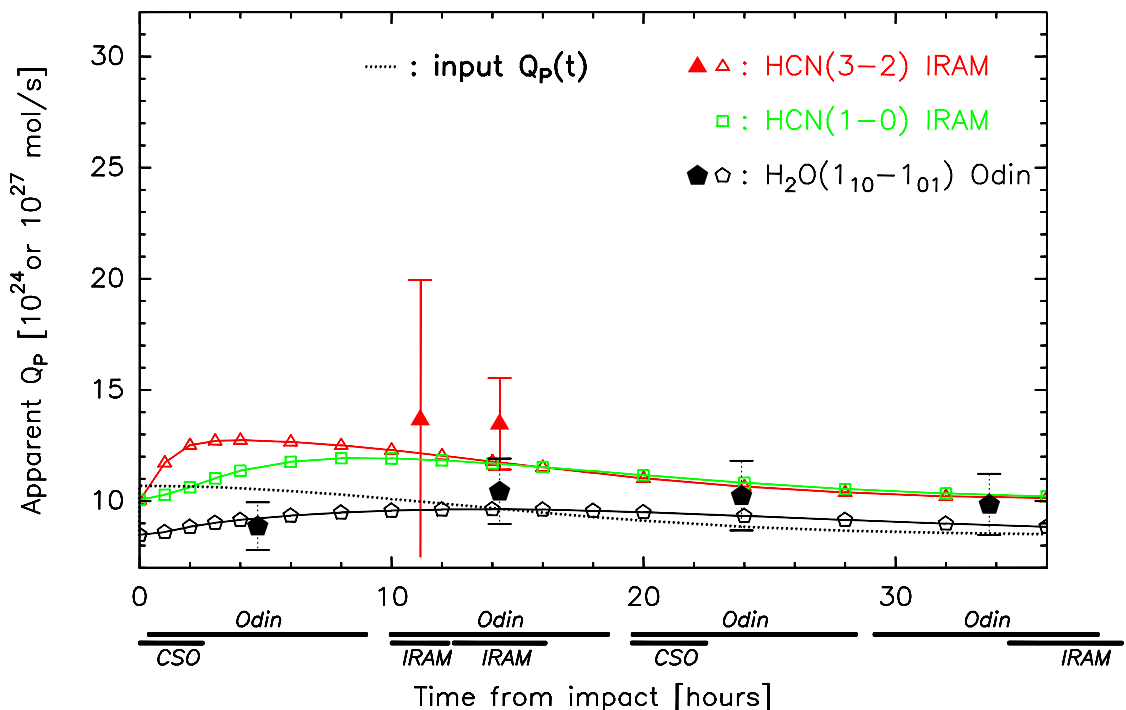}\vspace{-0.5cm}
   \caption{Same as in Fig.~\ref{figsimb1h} but for molecules released in 15~h.}
   \label{figsimb15h}
\end{figure*}

Two kinds of outbursts were considered. In the first case, we  
assumed that the outburst results in an isotropic
distribution of the material. Released HCN and CH$_3$OH molecules during the
outburst are in 0.12\% and 2.7\% relative proportion with respect to water, 
respectively. We investigated the cases  
$\Delta t=1$~h (i.e., $\Delta Q/Q=4.24$) and $\Delta t=15$~h 
($\Delta Q/Q=0.28$). The temperature and expansion velocity of the gas 
are supposed to remain constant with time and throughout the coma. 
For illustration purposes, the cases ($\Delta t=1$ and 15~h) for HCN and H$_2$O
are shown in Fig.~\ref{figsimb1h} and~\ref{figsimb15h}.

\bigskip
[Fig.~\ref{figsimb1h} and~\ref{figsimb15h}]
\bigskip

\begin{figure*}\vspace{-1cm}\hspace{1cm}
   \psfig{width=15cm,angle=270,figure=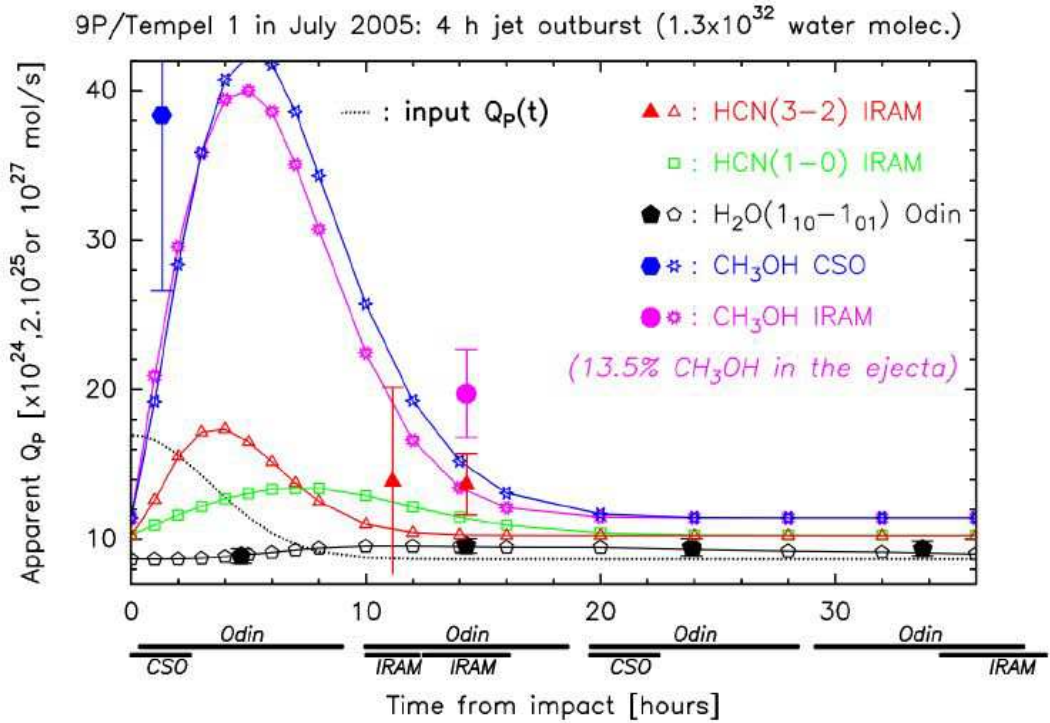}\vspace{-0.5cm}
   \caption{Simulation of the effect of an outburst of $1.3\times10^{32}$ water
	molecules released in 4~h (dotted line). The release is limited 
        to a jet in the plane of the sky with a $\pi/2$ 
	steradians opening angle and a velocity of 0.35~\kms.
        The HCN abundance relative to water is 0.12\%.    
	For CH$_3$OH, the abundance relative to water
	inside the jet is 13.5\%, versus 2.7\% in the surrounding coma,  
	as suggested by the observations.   
	Measurements ($8\times10^{27}$~\mols + outgassing in a jet from 
	Table~\ref{tabqh2oimp} for water; mean value plus excess found
	for the other molecules) are plotted with filled symbols and 
	error-bars.}
   \label{figsimbjet4h}\vspace{-0.5cm}
   \end{figure*}

\begin{figure*}\vspace{-1cm}\hspace{1cm}
   \psfig{width=15cm,angle=270,figure=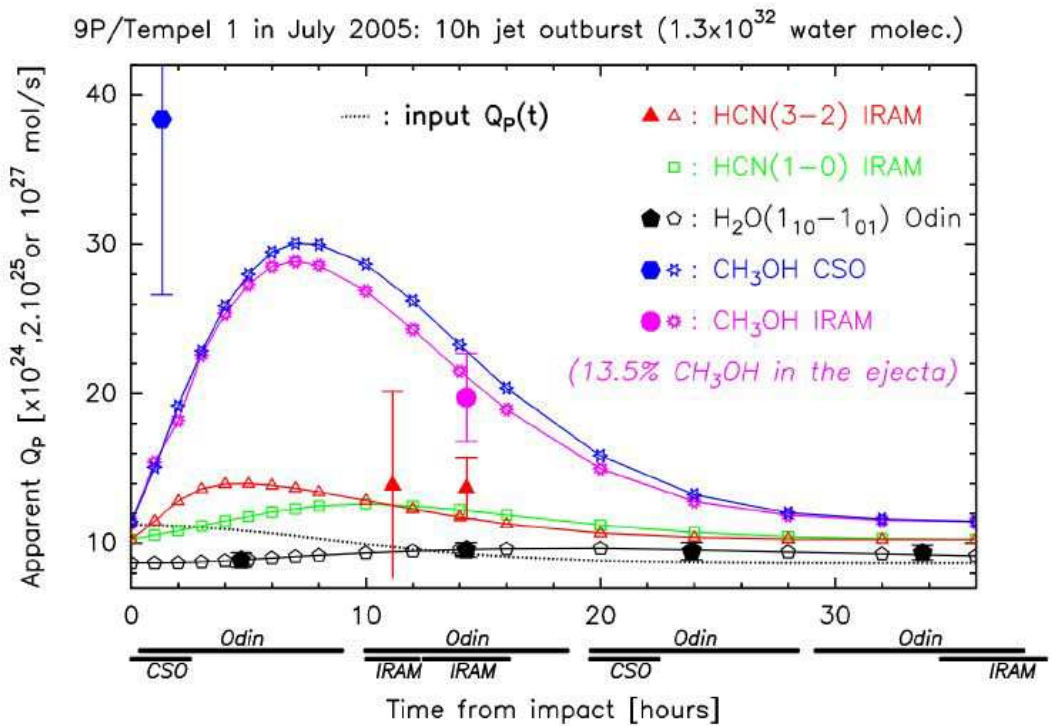}\vspace{-0.5cm}
   \caption{Same as in Fig.~\ref{figsimbjet4h} for a release lasting 10~h.}
   \label{figsimbjet10h}
\end{figure*}

In another simulation the outburst is restricted to a cone 
of $\pi/2$ steradians 
in the plane of the sky with molecules outflowing at a lower velocity 
($v_{exp}=0.35$~\kms). This more extreme case will provide 
a better simulation of the observed H$_2$O line shapes (Section 5.1),
and is more realistic since the centre of the cone of 
ejecta produced by the impact was indeed close to the
plane of the sky (A'Hearn et al. 2005b). We also assume that the abundance
of CH$_3$OH relative to water is 13.5\%, i.e., 5 times the value in the 
surrounding coma for direct comparison with the observations (Sect. 5.2).  
The simulations for $\Delta t=4$~h and 10~h are shown in
Figs.~\ref{figsimbjet4h} and~\ref{figsimbjet10h}.

\bigskip
[Figs.~\ref{figsimbjet4h} and~\ref{figsimbjet10h}]
\bigskip

\bigskip
[Table~\ref{tabsimul}]
\bigskip

Table~\ref{tabsimul} summarizes the
characteristics of the simulated bursts that would be observed for H$_2$O 
with the Odin beam and HCN~$J$(3--2) with the IRAM~30-m beam. In summary, 
the following results are found:

\begin{enumerate}
\item Isotropic burst with $\Delta t$ = 1~h: 
the outburst is seen with a delay of  
3--13~h for H$_2$O($1_{10}-1_{01}$), 1.5~h for HCN~$J$(3--2) and about 4~h
for HCN~$J$(1--0);

\item Isotropic burst with $\Delta t$ = 15~h: the outburst is seen with a 
delay of $\approx$14~h for H$_2$O($1_{10}-1_{01}$), $\approx3$~h for 
HCN~$J$(3--2) and about 7~h for HCN~$J$(1--0);

\item $\Delta t$ = 1 to 10~h in a jet of $\pi/2$ steradians in the plane of
the sky ($Q_{\rm H_2O}=10^{27}$~\mols in this cone before burst):
the outburst is seen with a delay of about 1 day for H$_2$O($1_{10}-1_{01}$)
and 3--4~h on average for HCN~$J$(3--2). The maximum is also significantly 
attenuated for the H$_2$O line for the short duration burst due to opacity 
(at the beginning the local density is multiplied by 56 if $\Delta t=1$~h), 
(Figs.~\ref{figsimbjet4h} and~\ref{figsimbjet10h});
\end{enumerate}

Because the time delay is 0.5 to 1~day for H$_2$O with Odin,  
continuous observations over nearly
two days after the impact (as performed) are needed to possibly detect 
impact-related water
excesses. On the other hand, a few (3--7) hours after the end of the
outburst, we do not expect significant impact-related signal excesses   
in IRAM and CSO observations. A more continuous worldwide coverage
would have been useful. 

The simulations also show that, under the assumption of an isotropic outburst,
the total number of molecules injected in the burst can be retrieved by 
integrating over time the apparent production rates. This is indeed expected
for optically thin lines. There are only limited photon losses
for H$_2$O (10\% for the 1~h burst, much less for the 15~h duration burst) 
due to moderate opacity effects. Losses become important for water 
when considering a
cone-restricted outburst, because of high local densities: 64, 32 and 18 \%
for 1, 4, 10~h bursts, respectively, for the geometry considered here.

\section{Intrinsic variations of the outgassing and molecular abundances}

Table~\ref{tabqprod} lists ``apparent'' production rates computed 
according to Section 3.1. In this section, we will present long-term 
trends and periodic variations observed in the data, a necessary step
to determine relative molecular abundances and study the increase of
molecular production due to the impact excavation.

\bigskip
[Table~\ref{tabqprod}]
\bigskip

\subsection{Long-term evolution of the outgassing}

\begin{figure*}\vspace{-0cm}\hspace{1cm}
   \psfig{width=15cm,angle=270,figure=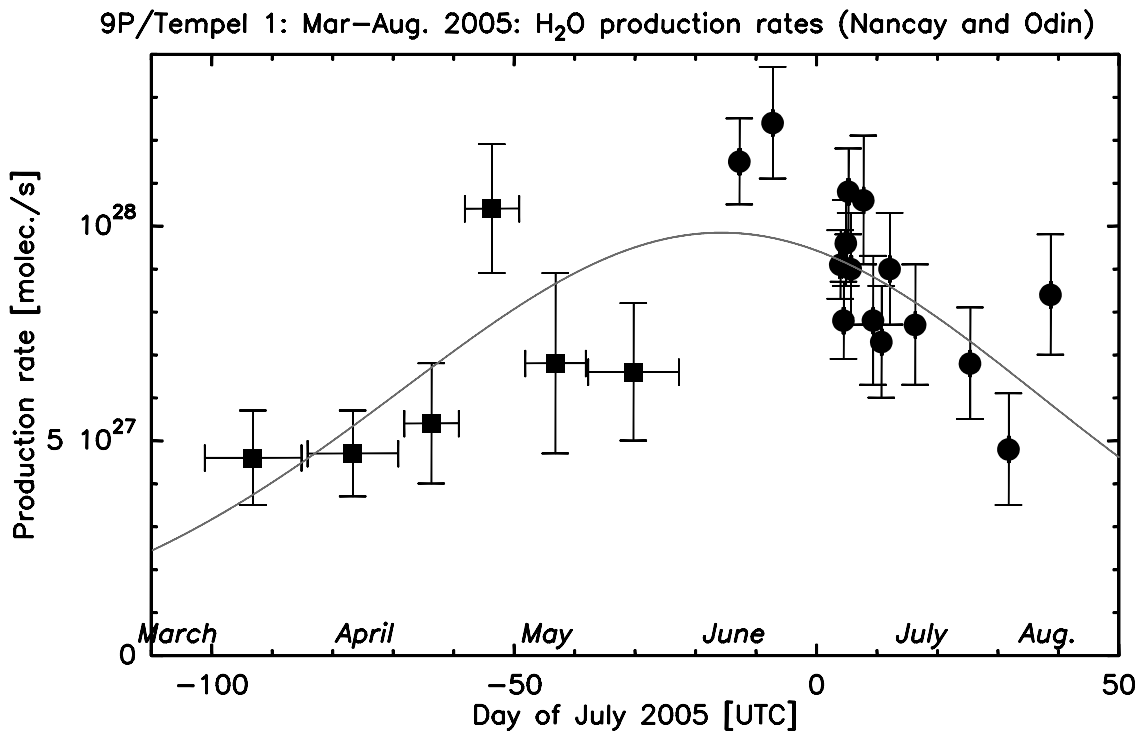}\vspace{-0.5cm}
   \caption{Water production rates based either on Nan\c{c}ay 
	observations of the OH radical (squares) or Odin observations 
	of the H$_2$O line at 557~GHz (circles). The continuous curve 
        corresponds to Eq.~(\ref{fqlongterm}) deduced from the least squares
        fitting.}
   \label{figqh2oall}
\end{figure*}

Fig.~\ref{figqh2oall} shows the evolution of the water production rate 
from March to August 2005.   
The least squares fitting of a Gaussian to the 
water production rates, taking into account a $r_h^{-2}$ heliocentric 
evolution due to increased solar irradiance as the comet gets close to the 
Sun, yields:
 
\begin{equation}
Q_{\rm H_2O} = 10.0\times10^{27}\exp\left(-\left(\frac{t+23}{88}\right)^2\right)\times\left(\frac{1.506}{r_h}\right)^2, 
\label{fqlongterm}
\end{equation}

\noindent
where $t = $date$-T_{perihelion}$ in days.
The 23 June data point is not included in this analysis (cf. Section 6). 
The fit is marginally significant ($\chi^2=27$ versus 41 for no variation), 
and might be biased by the inhomogeneity of the data (Nan\c{c}ay versus Odin).
Indeed, H$_2$O production rates derived from Nan\c{c}ay and Odin
simultaneous measurements show sometimes significant discrepancies 
(Colom et al. 2004). A peak of activity before perihelion is suggested. This 
is in agreement with previous perihelion passages, though the peak
in production was near 2 months before perihelion rather than 3 weeks (see the
review of Lisse et al. 2005). This pre/post perihelion asymmetry could be 
due to the presence of an active region close to the nucleus south pole 
which is no longer illuminated after May--June (a fan-shaped structure
was indeed observed to the south of the comet in March--May optical images).
This also suggests a mean production rate around the time of the impact of
$9.3\times10^{27}$~\mols, in agreement with other measurements: 
$6\times10^{27}$ (Schleicher et al. 2006), 
$7\times10^{27}$ (Bensch et al. 2006) 
to $10\times10^{27}$~\mols (Mumma et al. 2005).

\bigskip
[Fig.~\ref{figqh2oall}]
\bigskip

\subsection{Periodic variations of the HCN production rate}

\begin{figure*}\vspace{-0cm}\hspace{1cm}
   \psfig{width=15cm,angle=0,figure=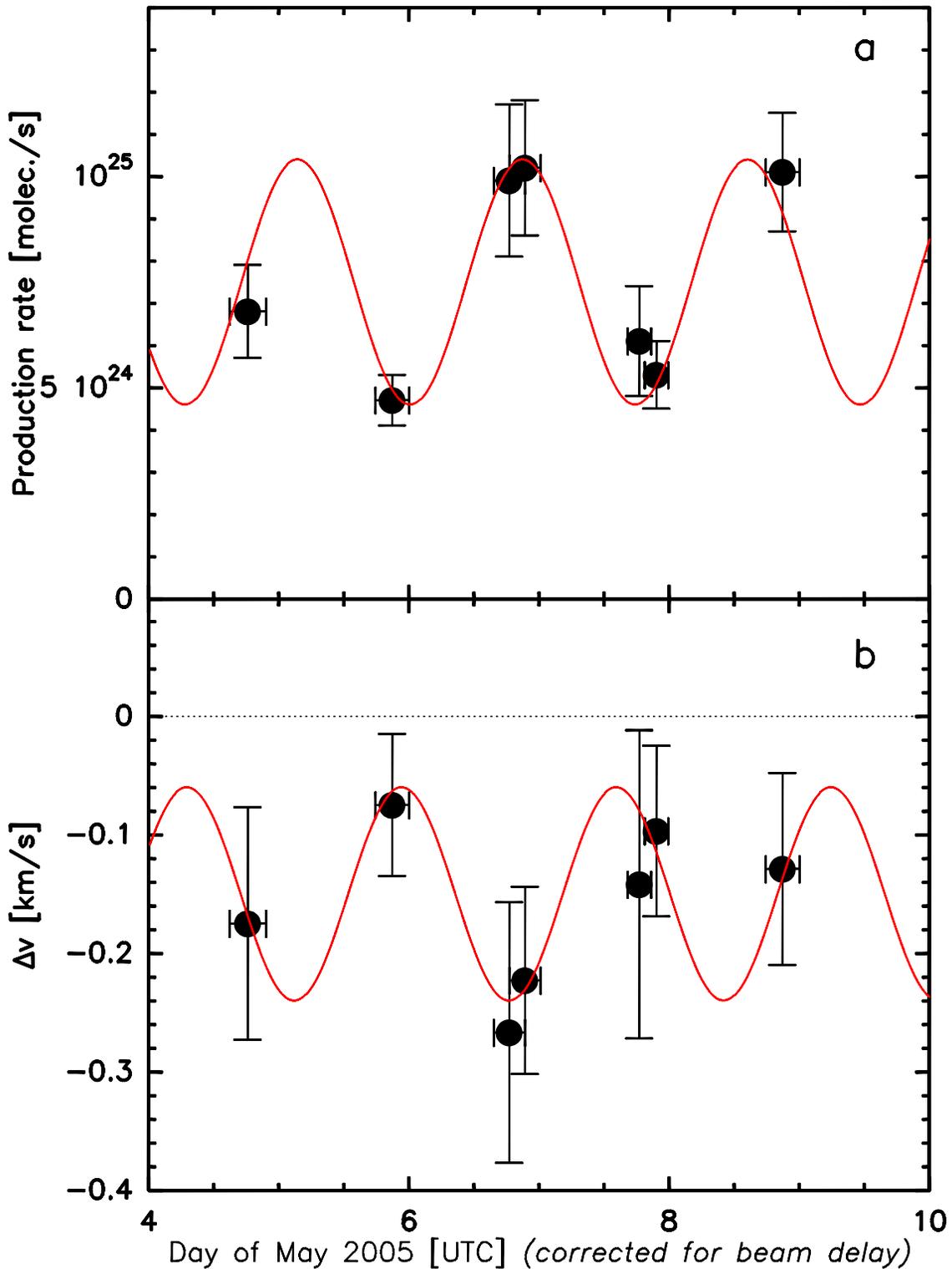}\vspace{-0.5cm}
   \caption{Top: HCN production rates based on IRAM~30-m observations 
	in May 2005, based on isotropic outgassing. 
	A sinusoidal fit to the data is plotted (Eq.~(\ref{fqhcn})).
	Bottom: Doppler shifts of the HCN$J$(1--0) and HCN$J$(3--2) lines 
	observed at IRAM~30-m in May 2005. 
	The sinusoidal fit to the data (Eq.~(\ref{fdvhcn})) is plotted.
	All times were corrected for ``beam delay'' (cf. text).}
   \label{figqpdvhcn}
   \end{figure*}

Fig.~\ref{figqpdvhcn}a shows the HCN ``apparent'' production rate versus time
observed in May. Data were split into two time intervals per night
and the production rates based on HCN(1--0) and HCN(3--2) plotted separately
(although the temperature is one fixed parameter that makes
them not fully independent). Day-to-day variations with some periodicity 
are observed, as for the line intensities (Table~\ref{tabobs}).

We first searched for periodicity with the phase dispersion minimization (PDM)
method (Stellingwerf 1978). This method is powerful in unveiling 
periodicities in a signal, especially when data are irregularly 
sampled and scarce (Colom and G\'erard 1988). 
The result of the PDM method applied to the 14 data
points shows a significant minimum of the variance ratio (around 0.23 
which means a significance level of 99\%) around 2 days with deeper peaks
at 1.8 and 2.3 days. The two periods are the result of an aliasing
($1/T_p' = 1/T_s-1/T_p$ where $T_s\approx1$~day is the 
the sampling period).
The most significant peak is close to the $\approx1.7$~day estimated 
rotation period of the nucleus (Belton et al. 2005, A'Hearn et al. 2005b).

\begin{figure*}\vspace{-0cm}\hspace{1cm}
   \psfig{width=15cm,angle=270,figure=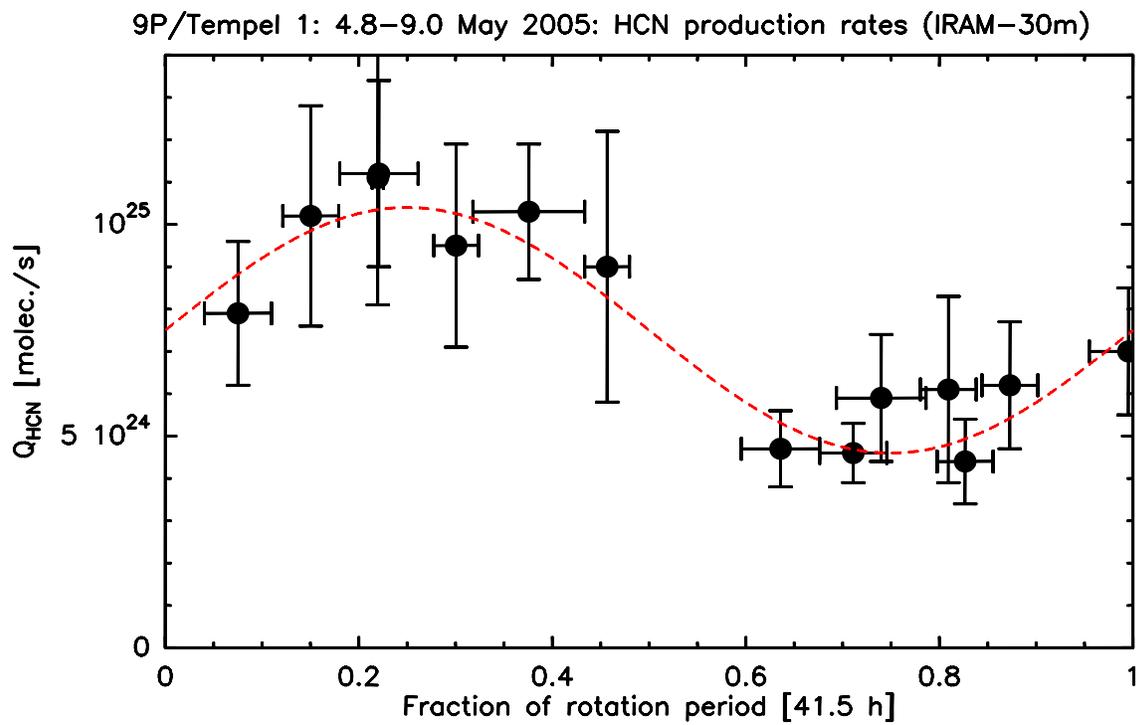}\vspace{-0.5cm}
   \caption{HCN production rates based on IRAM~30-m observations 
	in May 2005 folded on the 1.73 rotation period 
	(Fig.~\ref{figqpdvhcn}, Eq.~(\ref{fqhcn}). 
	All the data points (2 per day per line) used in the
	sine adjustment are shown here.
	All times were corrected for ``beam delay'' (cf. text).}
   \label{figqphcnfold}
   \end{figure*}

The second method was direct fitting of a sine curve to the data. 
A sine evolution was assumed since the signal-to-noise ratio is not high 
enough to determine the shape of a periodic variation.
 Indeed, folding all the 14 points 
(or 7 points if we average daily measurements as in 
Fig.~\ref{figqpdvhcn} to increase the signal-to-noise) over one period
cannot discriminate the variation of 
$Q_{\rm HCN}$ from a simple sine evolution (Fig.~\ref{figqphcnfold}).
A 4-parameter (period $T_p$, phase origin $T_0$, 
mean production rate $Q_{m\rm HCN}$ and amplitude 
$\Delta~Q_{\rm HCN}$) weighted least squares fitting over 14 points yields:
\begin{itemize} 
 \item $T_p=1.68$ days;
 \item $T_0=6.51$ May 2005;
 \item $Q_{m\rm HCN}=7.3\times10^{24}$~\mols;
 \item $\Delta~Q_{\rm HCN}=3.0\times10^{24}$~\mols;
 \item $\chi^2=3.8$ and the reduced chi-square is $\chi^2_{\nu=10}=0.38$.
\end{itemize}

\noindent
After ``beam delay'' correction, we get (Sect. 3.2.1):

\begin{itemize} 
 \item $T_p=1.73\pm0.10$ days;
 \item $T_0=6.44\pm0.07$ May 2005;
 \item $Q_{m\rm HCN}=7.5\pm0.5\times10^{24}$~\mols;
 \item $\Delta~Q_{\rm HCN}=2.9\pm0.7\times10^{24}$~\mols;
 \item $\chi^2=3.1$ and the reduced chi-square is $\chi^2_{\nu=10}=0.31$.
\end{itemize}
\noindent
which gives: 
\begin{equation}
Q_{\rm HCN}=\left(7.5+2.9\times\sin\left[2\pi\frac{t-6.44{\rm~May~2005}}{1.73}\right]\right)
\times10^{24}\mols.
\label{fqhcn}
\end{equation}
This curve is plotted in Fig.~\ref{figqpdvhcn}a.
The uncertainty on the fitted parameters is obtained by the projection of
the 4-D ellipsoid $\Delta\chi^2=1$ on the four parameters axes.

\bigskip
[Fig.~\ref{figqpdvhcn}]
\bigskip
\bigskip
[Fig.~\ref{figqphcnfold}]
\bigskip

The period of HCN variation is very close to the actual rotation period of the
nucleus (1.701 days, A'Hearn et al. 2005b). A HCN production curve 
with two peaks per period, as observed for the nucleus lightcurve, seems 
excluded from our data. This suggests that one side of the nucleus might have 
been more productive than the other when illuminated by the Sun. 
Jehin et al. (2006) 
found a periodic variation of 1.709 days in the CN and NH fluxes measured 
in June and around the impact date. Two major peaks were observed in one rotational
phase, which were interpreted by the presence of two active regions, the
strongest one corresponding to a peak of outgassing taking place on 
July 4.32 UT equivalent date.
A search for periodicity in the July HCN data, for direct 
comparison with CN data, is not possible given their poorer quality and low 
signal-to-noise ratios. If the strongest peak seen in CN by 
Jehin et al. (2006),
assuming CN mainly comes from the photodissociation of HCN,
corresponds to the maxima observed for HCN in May, then the observations
are compatible with a ``Tempel day'' of 1.719~days (34 rotations between 
May 6.88 and July 4.32) or $T_p=1.715$ days for the sideral rotation period.

The strong variation in HCN production (a factor of $\approx$3) 
during nucleus rotation shows that ``natural'' variability
of the comet activity must be taken into account when searching 
for impact-related production excesses.    

\subsection{Variation of line shapes}

Periodic variations are also present in the HCN line shapes observed in May.
Fig.~\ref{figqpdvhcn}b shows the evolution of the Doppler shift derived 
from daily averaged data (Table~\ref{tabobs}). The least squares 
sinusoidal fit adjustment yields : 

\begin{equation}
\Delta~v = -0.15+0.09\sin\left[2\pi\frac{t-7.18{\rm~May~2005}}{1.65}\right] ~\kms,
\label{fdvhcn}
\end{equation}
with the date $t$ in days corrected for ``beam delay''. 
The period of variation $T_p=1.65^{+0.31}_{-0.14}$ days is in agreement 
with previous findings. The reduced $\chi^2$ is $\chi^2_4=0.13$, 
versus $\chi^2_6=0.69$ for a straight line fit to the 7 data points 
with mean Doppler shift $\Delta v=-0.14$~\kms.

There is a correlation between line shapes and line intensities 
(Table~\ref{tabobs}, Fig.~\ref{fig9phcnjet}). The blueshift of the lines is 
at its maximum at the time of maximum outgassing. This can be explained by
an active source which molecular production varies with solar illumination 
when the nucleus is rotating. A simple simulation with a broad ($\pi/2$ 
steradians) rotating HCN jet actually provides a good match to the observed 
line strengths and shapes.

\subsection{Short-term variation of the water production rate}

\begin{figure*}\vspace{4cm}\hspace{1cm}
   \psfig{width=15cm,angle=270,figure=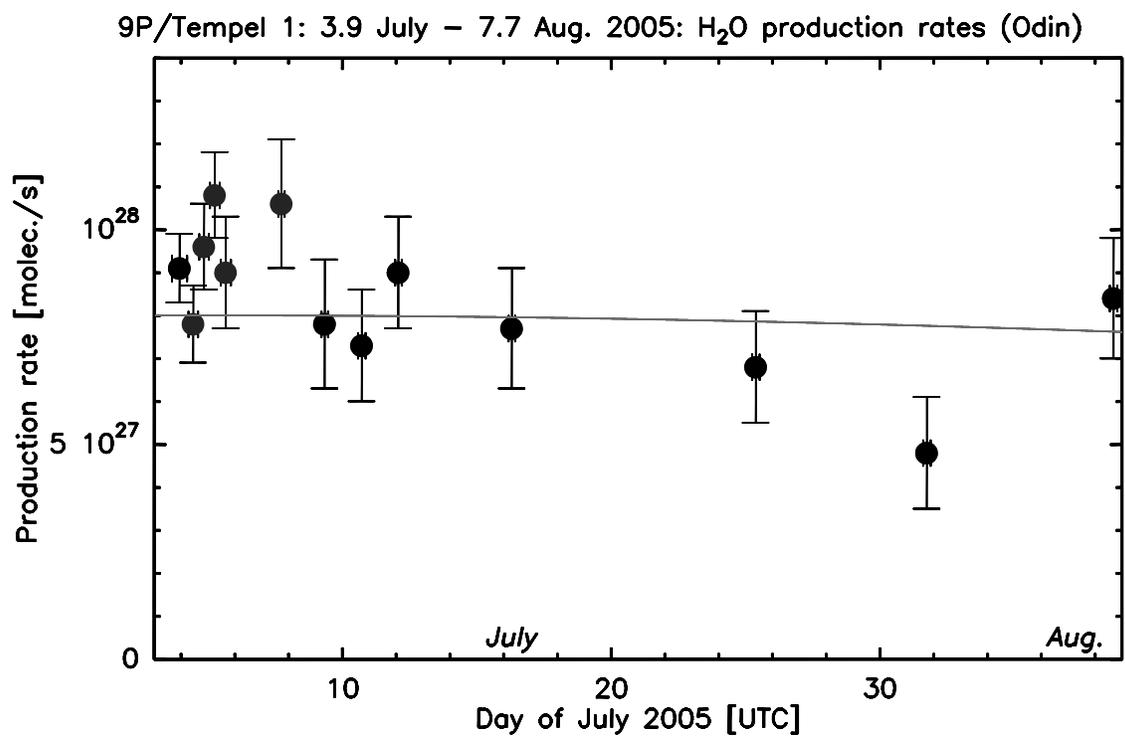}\vspace{-0.5cm}
   \caption{Apparent H$_2$O production rates measured from Odin observations 
    and the least squares fit of $Q/r_h^2$ to the data.}
   \label{figqh2o}
\end{figure*}

The evidence of periodic variation in outgassing seen in HCN May data
or for several gaseous species in June--July (Jehin et al. 2006) 
convinced us to look for variation in H$_2$O production rates measured
with Odin. But as explained in Sect.~3.2.1, we expect the amplitude
of the variation of the H$_2$O($1_{10}-1_{01}$) line observed by Odin
to be 3.5 times lower than that for HCN observed at IRAM.
In Fig.~\ref{figqh2o} we have plotted the average $Q_{\rm H_2O}$
corrected for $1/r_h^2$ heliocentric evolution and excluding the 
July 4.4--7.8 post-impact data.
The mean value is $Q_{m\rm H_2O}=8.0\times10^{27}$~\mols at perihelion,
so that the apparent amplitude of periodic variation similar to those seen 
for HCN would be on the order of $0.9\times10^{27}$~\mols, similar to
each individual measurement uncertainty.

Actually fitting a sine evolution with fixed phase 
($T_0=3.90$ July + 0.82 day (cf. Sect.~4.2 and ``Beam delay'') and 
period (1.71 days), yields:
\begin{equation}
Q_{\rm H_2O}=
\left(8.1+0.7\times\sin\left[2\pi\frac{t-4.72}{1.71}\right]\right)
\times\left(\frac{1.506}{r_h}\right)^2\times10^{27}\mols.
\label{fqodin}
\end{equation}
The reduced chi-square is $\chi^2_6=1.32$, to be compared to  
$\chi^2_5=1.29$ when fitting just the mean value 
$Q_{m\rm H_2O}\times(1.506/r_h)^2$. Periodic variations cannot
be reliably retrieved from Odin data, but the attempt to fit a sine variation
to the data shows that a strong variation of outgassing with
rotation phase of the nucleus is not excluded: the fitted amplitude 
is similar to the expected one (Sect.~3.2.1), about 1/3${\rm rd}$ of the 
amplitude of the CN variations seen by Jehin et al. (2006) to 
1/4$^{\rm th} of Q${\rm HCN} variations in May. 
So, a variation in natural outgassing should be taken into account when 
looking for impact related effects. Indeed, as shown by Jehin et al. (2006), 
$T_0\approx3.90$ July means that ``natural'' outgassing of the nucleus
was in its rising phase at the impact time (4.244 July), peaking 3--4~h later.

\bigskip
[Fig.~\ref{figqh2o}]
\bigskip

For the purpose of determining molecular abundances based on IRAM 
observations in May, we will use a reference water production rate:

\begin{equation}
Q_{\rm H_2O,ref}=
\left(8+3\sin\left[2\pi\frac{t-0.1-({\rm 6.44~May~2005})}{1.72}\right]\right)
\times\left(\frac{1.506}{r_h}\right)^2\times10^{27}\mols.
\label{fqrefiram}
\end{equation}

It matches the May periodicity of HCN
and the April-May mean water production rate derived from Nan\c{c}ay 
observations ($7\pm1\times10^{27}$~\mols). 
$t$ is the number of days since 0.0 May 2005 UT with
a 0.1 day shift added to take into account the ``beam delay'' for 
millimetre observations. The computed reference water production rates 
are given in column 5 of Table~\ref{tabqprod}.  

\subsection{Mean molecular abundances}
	Molecular production rates were ratioed to the reference water
production rate in order to derive the molecular abundances given
in Table~\ref{tababund}. The abundances did not vary significantly between
May and July and the mean values (H$_2$O:CO:CH$_3$OH:H$_2$CO:H$_2$S:HCN:CS) = 
(100~:~$<10$~:~2.7~:~$<$1.5~:~0.5~:~0.12~:~$<$0.13) fall well within the 
typical abundances measured in about 30 comets since 1986 
(Biver et al. 2002, 2006a).

\section{Amount of gas released by the impact}
	Cautiously taking into account intrinsic periodic 
variations of the outgassing rate, we can now estimate the   
contributions from the impact or outburst related activity. 
However, one problem is that the excess of signal will
not only depend on the number of molecules released but also
on their velocity and the duration of the burst in outgassing.
The simplified simulations presented in Sect.~3.2 will guide our 
analysis.

\subsection{Water}
	For H$_2$O data, the simplest way (model (1)) to evaluate the 
outgassing excess due to the impact is to compute the difference 
between the observed apparent production rates (Table~\ref{tabqh2opha}, 
column 4) and a reference production rate.
The reference production rates are taken from the last column of 
 Table~\ref{tabqh2opha}. 
The right part of this table provides lines characteristics and
apparent production rates based on Odin observations between
7 July and 7 August binned according to the expected rotation phase
(Column 1).
Using the sine function of Eq.~\ref{fqodin} would give very similar results.

\bigskip
[Table~\ref{tabqh2opha}]
\bigskip

\begin{figure*}\vspace{4cm}\hspace{1cm}
   \psfig{width=15cm,angle=270,figure=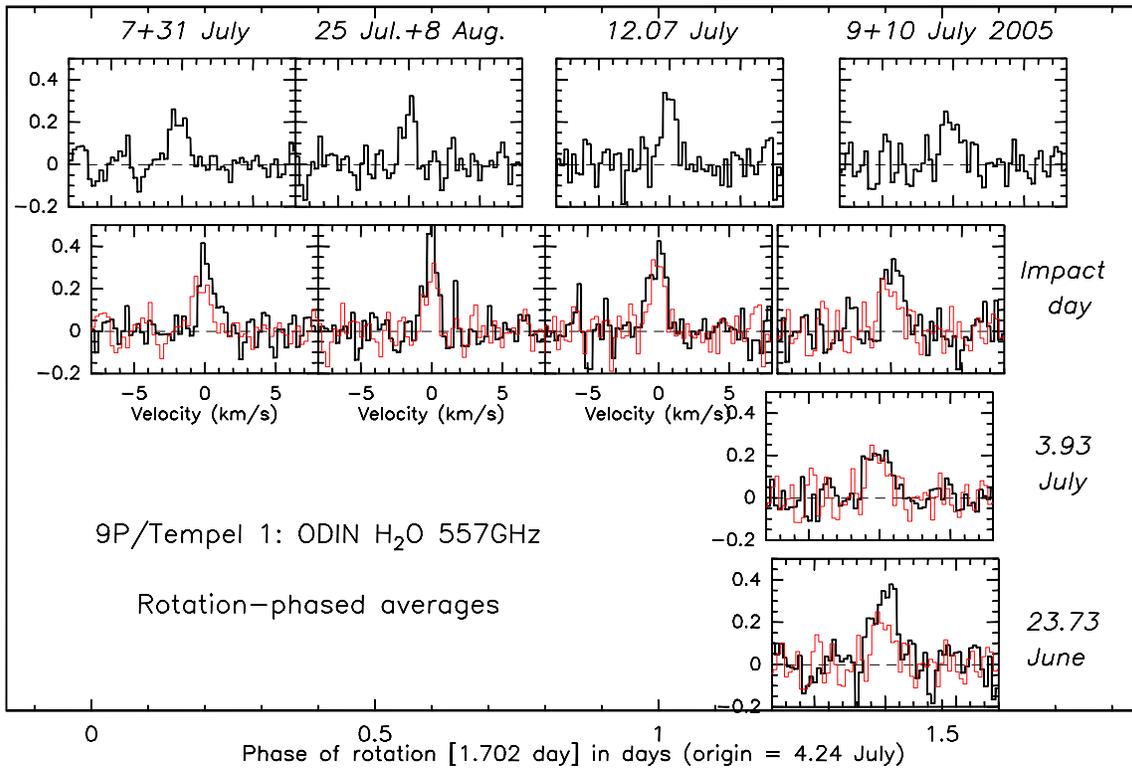}\vspace{-0.5cm}
   \caption{Spectra of H$_2$O obtained with Odin summed and ordered
	according to the phase of the comet rotation (1.7 day, horizontal 
	scale) and time (vertical scale, from before the impact (bottom) to 
	after the impact (top row)).}
   \label{fig9ph2opha}
\end{figure*}

\begin{figure*}\vspace{-0.0cm}\hspace{1cm}
   \psfig{width=14cm,angle=270,figure=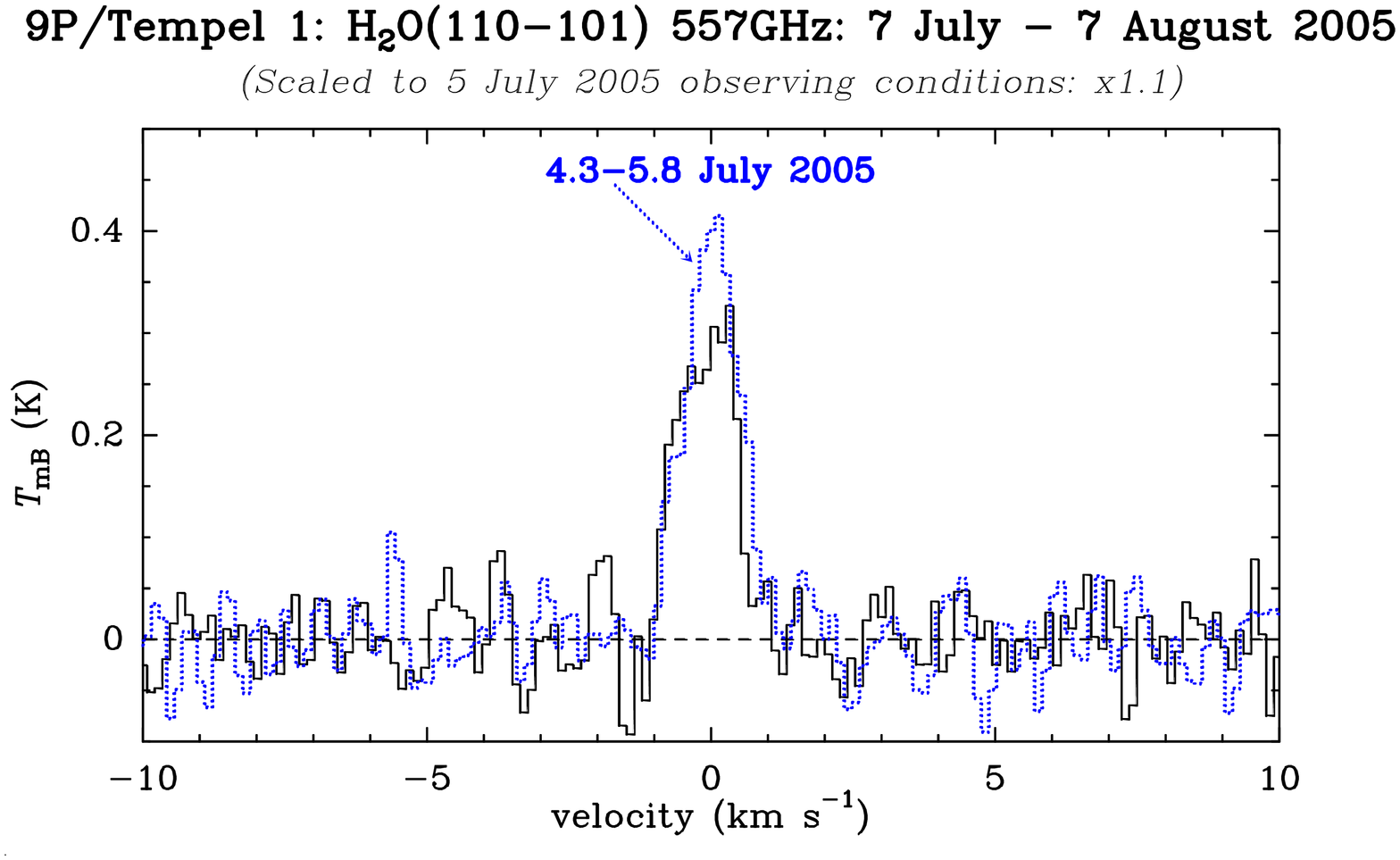}\vspace{-0.5cm}
   \caption{The 557 GHz H$_2$O line observed by Odin
        after the Deep Impact collision. Dotted line: an average of observations 
        obtained during the first rotation just following the impact.
        Solid line: an average of observations obtained between 7.7 July and
        7.7 August, after dissipation of the impact ejecta cloud. These observations
        were sampled and averaged to cover a full nucleus rotation. They 
        were scaled to correct for the decrease of the signal 
        due to the change in geocentric and heliocentric distances.}
   \label{fig9ph2osum}
\end{figure*}

The main difference between
the reference and impact spectra is a narrow spike around zero 
velocity which is superimposed on a broader component and is only seen
on the 4.4--5.6 July high resolution spectra (Figs.~\ref{fig9ph2opha} and 
\ref{fig9ph2osum}). Correcting for the slightly different geometries
at different dates, the line integrated intensity of this spike was 
estimated, as given in Table~\ref{tabqh2oimp}. It is close to the excess
signal seen above $T_{mb} = 0.25$ K. 
The conversion of this line intensity into a number of molecules is not 
straightforward since the H$_2$O line is optically thick: 
photons emitted by the cloud of ejecta responsible for this spike will be 
absorbed by the foreground H$_2$O coma. This process was not considered.   
We expect that the results are not affected by a large factor, since both   
the spike position ($v\approx 0.0$ \kms) and geometry of the ejecta 
suggest that this cloud was in the plane of the sky, minimizing
line-of-sight opacity effects. The  narrow width of the spike 
($HWHM=0.22\pm0.05$~\kms) also suggests 
a relatively low velocity of $v_{exp}\approx0.35$~\kms, if we assume an 
opening angle of 90\deg~for the cloud of ejecta. This low velocity  
also minimizes photon absorption by the higher 
velocity surrounding gas. For this second estimate of the amount of gas 
inside the ejecta (model (2)), we run our radiative transfer code 
with the molecules distributed inside a cone of opening angle of 90\deg, and
expanding at $0.35$~\kms.   

\bigskip
[Figs.~\ref{fig9ph2opha},~\ref{fig9ph2osum}]
\bigskip

Table~\ref{tabqh2oimp} provides 
impact-related apparent production rates as a function of time, 
and corresponding total amounts of H$_2$O molecules, for the two 
different analyses. 
As discussed in Sect. 3.2, these numbers 
should be close to the total number of molecules 
released in the burst. 

\bigskip
[Table~\ref{tabqh2oimp}]
\bigskip

Although based on the same data, models (1) and (2) give different results for
the apparent production rate (line 5 in Table~\ref{tabqh2oimp}). This is
because model (2) assumes a lower velocity in order to interpret the narrow 
spike. However, similar amounts of molecules are derived from the two models.
Indeed, for model (2) we take into account that part 
of the emitted photons are lost by self-absorption effects, as discussed
in Sect. 3.2. The time evolution of the 
apparent production rates derived from 
model (2), when compared to the simulations, provides a good agreement 
with a 4--10~h duration outburst (Figs.~\ref{figsimbjet4h} 
and ~\ref{figsimbjet10h}) and $N_t=1.3-2\times10^{32}$ 
molecules. Mumma et al. (2005) found an excess of $10^{28}$~\mols in 
water production in the 1--3~h following the impact, on the basis of infrared
measurements inside a $\approx1$\arcsec aperture for which beam dilution 
and beam delay can be neglected. The sublimation of 5000~tons of water ice
at a rate of $\approx10^{28}$~\mols would take $\approx$ 4~h. Therefore, 
the observations of Mumma et al. (2005) are consistent with the simulation 
shown in Fig~\ref{figsimbjet4h}. 

In summary, the two approaches yield consistent values for the 
amount of water released by the impact. 
We adopt a mean value of $5000\pm2000$ tons ($N_t=1.7\times10^{32}$ 
molecules), given that the increase of outgassing due to the impact is 
detected at the 3--4 $\sigma$ level 
(line 5, column 7 of Table~\ref{tabqh2oimp} and Fig.~\ref{fig9ph2osum})
and that the uncertainty due to excitation parameters (e.g. electron density)
can be on the order of 10\% or slightly higher.
This amount is also consistent with estimates based on near-UV observations
of OH by K\"uppers et al. (2005) ($\approx4500$ tons) and 
Schleicher et al. (2006) ($<13000$ tons).

\subsection{Other molecules}
	
Due to limited signal-to-noise ratios ($\approx5$ for HCN and CH$_3$OH,
versus 10 for H$_2$O), the millimetre lines do not show clear evidence of
emission excesses related to the impact. Most of the 
observations took place more than 10~h after the impact, which  
implies that most of the material had already left the beams. 
From the various 
simulations discussed in Section 3.2, we can mainly note that the beam 
delays are on the order of 2~h (HCN~$J$(3--2)), 4~h (CH$_3$OH) to 7~h 
(HCN~$J$(1--0)).
Therefore the millimetre molecular observations are much more sensitive 
to the time it took to the ejected 
icy grains to sublimate than are Odin H$_2$O observations.

Figs.~\ref{fighcn1} and ~\ref{fighcn3} show the spectra of both 
the HCN~$J$(1--0) and HCN~$J$(3--2) lines at IRAM on the day
of the impact (4.84 July 2005) and the averages before (2.8--3.8 July) and 
after (5.8 July) the impact. The lines appear to be stronger on 4.84 July 
(Table~\ref{tabobs}), which can be partly due to 
a larger intrinsic outgassing at that time (peak around 4.4 July). 
From May, 2--3 and 5--10 July data, we derive a mean abundance 
$Q_{\rm HCN}/Q_{\rm H_2O}=0.12\pm0.03$\% (Table~\ref{tababund}).    
The excess observed at $t_0$+14~h on 4.84 July is then:
\begin{equation}
\Delta~Q_{\rm HCN} = (13.1-0.0012\times8000)\pm2.1\times10^{24}\mols = 
3.5\pm2.1\times10^{24}\mols.
\label{fdqhcn}
\end{equation}
\noindent
No significant impact-related excess of HCN is thus detected.

The average abundance of methanol relative to water is 
$2.7\pm0.3$\% (Table~\ref{tababund}). Using this value as a
reference, the excess of methanol on 4.30 July, just after the impact
(assuming a background normal $Q_{\rm H_2O}\approx11\times10^{27}$~\mols, 
close to its possible peak value at that time) is:
\begin{equation}
\Delta~Q_{\rm CH_3OH}=5.5\pm2.4\times10^{26}\mols~~(t_0+{\rm1.3~h, CSO~data)}
\label{fdqch3ohcso}
\end{equation}
and on 4.84 July:
\begin{equation}
\Delta~Q_{\rm CH_3OH}=1.7\pm0.6\times10^{26}\mols~~(t_0+{\rm14~h, IRAM~data).}
\label{fdqch3ohiram}
\end{equation}
The methanol excess is still marginal however (2--2.5$\sigma$).

These results are too marginal to derive any secure conclusion.
We can only work on the hypothesis that 5000~tons of water were released
in probably 4~hours or more. According to Beer et al. (2006), 1--10$\mu$m
icy grains would sublimate in 0.5~h to more than 24~h at 1.5 AU,
depending on their water ice fraction. Deep Impact observations and 
XMM-Newton observations (Schulz et al. 2006) actually detected icy grains 
in the coma of the comet at least during the first hour following the impact.

The excesses of outgassing found for HCN and CH$_3$OH are  
plotted in Figs.~\ref{figsimbjet4h} 
and~\ref{figsimbjet10h} which present the evolution of the ``apparent'' 
production rate expected from the simulations for H$_2$O:CH$_3$OH:HCN ratios 
of 100:2.7:0.12. The measurements are compatible
with a normal abundance of HCN in the ejecta, while an overabundance 
of CH$_3$OH is suggested. 
Indeed, to explain both CSO and IRAM observations with the same
total quantity $N_{t\rm CH_3OH}$ of methanol released in the ejecta,
an outburst lasting around 4~h is required, with 
$N_{t\rm CH_3OH}=46\pm18\times10^{30}$ molecules. 
This implies a relative abundance CH$_3$OH/H$_2$O$ = 27\pm11$\% in the 
ejecta, i.e., one order of magnitude larger than the abundance measured
for normal activity. However, the methanol impact spectra  
(Figs.~\ref{figch3oh} 
and~\ref{figch3ohcso}) do not show any obvious feature such as the 
H$_2$O spike. So, it is not excluded that part of the excess signal 
(or loss on the other dates) is due to calibration or modelling 
uncertainties. Indeed
Mumma et al. (2005) find that the CH$_3$OH/H$_2$O ratio in the ejecta
was not significantly different from the background.  

Only upper limits have been obtained for molecules observed  
far from the impact time. An upper limit of $22$\% relative to
water is measured for CO on the day of the impact. Even assuming that
the CO abundance relative to water was 10\% (maximum allowed by
May observations) before the impact, the remaining 12\% yield an 
upper limit on the CO quantity in the ejecta much larger 
than that of water. This result does not provide useful constraints.

\section{Other outbursts of activity}
	Several bursts of activity were reported by the cameras of the
Deep Impact spacecraft before the encounter (A'Hearn et al. 2005b), or by other
observations in visible to X-ray wavelengths 
(e.g. on 8 July, Meech et al. 2005b).
Comparing the observed production rates of water with the expected 
natural variations, we found significant differences on two dates with  
$\Delta Q_{\rm H_2O}=4.0\pm1.2\times10^{27}$~\mols on 23.7 June and 
$\Delta Q_{\rm H_2O}=2.9\pm1.4\times10^{27}$~\mols on 7.7 July 
(Table~\ref{tabqprod}).
The stationary regime assumption is of course not relevant to analyze
these outbursts, and the $\Delta Q_{\rm H_2O}$
value given above (about +50\%) are just indicative.

The July 7.7 outburst does not correspond to a very strong or long outburst. 
Indeed, given the $\approx$ 0.6~day phase delay with IRAM observations, 
the Odin H$_2$O 
data can be compared to HCN or CH$_3$OH observations on 6.9 and 7.8 July, but 
nothing significant is observed on these dates. 
For a half-day burst (cf. Section 4.1), the $\Delta Q_{\rm H_2O}$ given above 
corresponds to 
$N_t\approx2\times10^{32}$ molecules, i.e., 1/3 day of normal activity. We
note, however, that this outburst is only detected at a 2-$\sigma$ level.

The Deep Impact team reported an outburst beginning on 22.38 June UT. The 
excess in H$_2$O emission that we observed 32~hours later is likely 
related to this outburst. In the spectrum, the extra signal due to 
this burst of
molecules is broader than the impact-related spike (Fig.~\ref{fig9ph2opha}). 
From a simulation similar to the one shown in Fig.~\ref{figsimb15h}, 
we infer a total release of $N_t\approx10^{33}$ molecules 
corresponding to 1.4 day of normal activity. This natural outburst 
surpassed significantly the one created by Deep Impact.

\section{Conclusion}


Comet 9P/Tempel 1 is certainly one of the weakest comets extensively
studied at radio wavelengths, with a peak outgassing rate around
$10^{28}$~\mols. This investigation campaign puts into evidence
several characteristics of the comet:

\begin{itemize}
	\item The relative molecular abundances are ``classical'', 
	of the order of (H$_2$O:CO:CH$_3$OH: H$_2$CO:H$_2$S:HCN:CS) =  
	(100:$<$10:2.7:$<$1.5:0.5:0.12:$<$0.13), 
	comparable to mean values observed in many comets.
	
	\item A strong regular variation of the outgassing rate is
	clearly observed in HCN data obtained in May 2005 
	($T_p=1.73\pm0.10$ days). Its amplitude is a factor of 3 from 
	minimum to maximum and its periodicity is likely related to a 
	rotation period.  Outgassing
	anisotropy is also evident.  This suggests that one side of
	the nucleus was more active than the other when illuminated.
	
	\item The periodic variation of the outgassing of the comet
	must be carefully taken into account when analyzing the
	effects of the Deep Impact collision. It is not excluded 
	that ``natural'' outgassing was still varying by $\pm40$\% 
	in early July and that it was in a rising phase at the time
	of the impact. 
  
	\item As regards to Deep Impact consequences, the total amount
	of water released is estimated to $5000\pm2000$ tons,
	corresponding to the cumulated production of 0.2 day of
	normal activity.  This water was most likely released in 
	icy grains that took several hours to sublimate: our best
	guess, based on simulations, is around 4~h.  It was
	not possible to assess precisely the composition of the
	ejecta, except for a possible increase in the abundance of
	CH$_3$OH; no large excess of HCN was seen.
	
	\item The comet also underwent significant natural outbursts
	of outgassing, possibly of larger amplitude than the
	Deep Impact related burst. One took place on 22--23 June
	and may have released as much water as 1.4~day of normal
	activity.
	
\end{itemize}

Although the signal was too weak to make a detailed
compositional study of the impact ejecta, this observing campaign
allowed us to observe in detail the behavior of a Jupiter-family
comet, and to put into evidence periodicity and outbursts of
outgassing poorly observed in any comet before.

\bigskip


{\it Acknowledgments.} 	
Odin is a Swedish-led satellite project funded jointly 
by the Swedish National Space Board (SNSB), the Canadian Space 
Agency (CSA), the National Technology Agency of Finland (Tekes) 
and the Centre National d'\'Etudes Spatiales (CNES, France). 
The Swedish Space Corporation is the
prime contractor, also responsible for Odin operations.
The authors thanks the whole Odin team, 
including the engineers who have been very supportive to
the difficult comet observations. Their help was essential in solving 
in near real time problems for such time critical observations. 
We are grateful to the IRAM and CSO staff and 
to other observers for their assistance during the observations. 
IRAM is an international institute co-funded by the 
Centre national de la recherche scientifique (CNRS), the Max Planck
Gesellschaft and the Instituto Geogr\'afico Nacional, Spain.
The CSO is supported by National Science Foundation grant AST-0540882. 
The Nan\c{c}ay radio observatory is the unit\'e scientifique de Nan\c{c}ay
of the Observatoire de Paris, associated as Unit\'e de Service et de 
Recherche (USR) 704 of the CNRS
The Nan\c{c}ay observatory also gratefully acknowledges the 
financial support of the Conseil r\'egional of the R\'egion Centre in France.

\newpage

\begin{table}\hspace{-1cm}
{\footnotesize
\begin{center}
\caption[]{Observations of OH 18-cm lines in comet 9P/Tempel 1 at Nan\c{c}ay}\label{tabobsoh}
\begin{tabular}{lccrllll}
\hline
UT date 2005 & $<r_{h}>$ & $<\Delta>$ &   Int. time & $\int T_bdv$    & Velocity offset & 
\multicolumn{2}{c}{$inversion$} \\[0cm]
 [mm/dd.dd-dd.dd] &  [AU] &  [AU]  &  [days$\times$1~h]  & [K~km~s$^{-1}$] & [km~s$^{-1}$]   & (1) & (2) \\
\hline
03/04.11--03/19.06 & 1.89 & 0.98 & 15 &  $+0.004\pm0.003$ &	&   -0.23 & -0.27 \\
03/20.06--04/05.01 & 1.78 & 0.82 & 12 &  $-0.015\pm0.003$ & $-0.70\pm0.69$	&   -0.28 & -0.33\\
04/06.01--04/21.95 & 1.71 & 0.75 & 17 &  $-0.013\pm0.002$ & $-0.29\pm0.16$	&   -0.31 & -0.37\\
04/22.95--05/01.92 & 1.65 & 0.72 & 10 &  $-0.014\pm0.003$ & $-0.08\pm0.14$	&   -0.30 & -0.36\\
05/02.92--05/11.89 & 1.62 & 0.71 &  9 &  $-0.029\pm0.004$ & $-0.23\pm0.14$	&   -0.28 & -0.34\\
05/12.89--05/22.86 & 1.58 & 0.72 & 10 &  $-0.015\pm0.004$ & $+0.13\pm0.20$	&   -0.25 & -0.30\\
05/24.86--06/08.82 & 1.54 & 0.76 & 14 &  $-0.011\pm0.002$ & $-0.41\pm0.17$	&   -0.19 & -0.21\\
07/01.78--07/10.77 & 1.51 & 0.90 & 10 &  $+0.011\pm0.004$ &	&   +0.03 & +0.08 \\
\hline
\end{tabular}
\end{center}
Note: Maser inversion values are from (1) model of Despois et al. (1981) or (2) Schleicher and A'Hearn (1988). 
}
\end{table}

\begin{table}
{\scriptsize
\begin{center}
\caption[]{Observations of comet 9P/Tempel 1}\label{tabobs}
\begin{tabular}{lccrclll}
\hline
UT date 2005 & $<r_{h}>$ & $<\Delta>$ &   Int. time &Species (transition)& $\int T_bdv$    & Velocity offset & Offset \\[0cm]
 [mm/dd.dd-dd.dd] &  [AU] &  [AU]  &      [min]  &      & [K~km~s$^{-1}$] & [km~s$^{-1}$]   &  \\
\hline
\multicolumn{8}{c}{Odin~1.1-m} \\
\hline
06/18.06--18.43	& 1.516 & 0.818 &  70 & H$_2$O($1_{10}-1_{01}$) & $0.67\pm0.10$ & $-0.31\pm0.13$ & 14\arcsec \\
		&	&	&  90 & H$_2$O($1_{10}-1_{01}$) & $0.37\pm0.09$ & $-0.08\pm0.25$ & 63\arcsec \\
		&	&	&  68 & H$_2$O($1_{10}-1_{01}$) & $0.33\pm0.10$ & $+0.38\pm0.32$ & 83\arcsec \\
06/23.61--23.85	& 1.511 & 0.842 & 179 & H$_2$O($1_{10}-1_{01}$) & $0.66\pm0.07$ & $-0.03\pm0.11$ & 14\arcsec \\
07/03.65--04.22	& 1.506 & 0.892 & 375 & H$_2$O($1_{10}-1_{01}$) & $0.44\pm0.04$ & $-0.14\pm0.08$ & 15\arcsec \\
07/04.26--04.62	& 1.506 & 0.895 & 248 & H$_2$O($1_{10}-1_{01}$) & $0.38\pm0.04$ & $+0.15\pm0.10$ & 14\arcsec \\
07/04.66--05.02	& 1.506 & 0.897 & 250 & H$_2$O($1_{10}-1_{01}$) & $0.46\pm0.05$ & $-0.09\pm0.10$ & 14\arcsec \\
07/05.06--05.43	& 1.506 & 0.899 & 250 & H$_2$O($1_{10}-1_{01}$) & $0.52\pm0.05$ & $-0.09\pm0.07$ & 15\arcsec \\
07/05.46--05.83	& 1.506 & 0.901 & 252 & H$_2$O($1_{10}-1_{01}$) & $0.41\pm0.05$ & $+0.17\pm0.13$ & 14\arcsec \\
07/07.57--07.84	& 1.506 & 0.912 & 218 & H$_2$O($1_{10}-1_{01}$) & $0.50\pm0.07$ & $-0.09\pm0.13$ & 12\arcsec \\
07/09.22--09.45	& 1.507 & 0.921 & 165 & H$_2$O($1_{10}-1_{01}$) & $0.37\pm0.07$ & $-0.06\pm0.19$ & 12\arcsec \\
07/10.57--10.85	& 1.507 & 0.929 & 167 & H$_2$O($1_{10}-1_{01}$) & $0.34\pm0.06$ & $-0.11\pm0.15$ & 15\arcsec \\
07/11.95--12.19	& 1.508 & 0.937 & 166 & H$_2$O($1_{10}-1_{01}$) & $0.42\pm0.06$ & $-0.21\pm0.16$ & 12\arcsec \\
07/16.17--16.41	& 1.510 & 0.962 & 206 & H$_2$O($1_{10}-1_{01}$) & $0.34\pm0.06$ & $-0.30\pm0.21$ & 18\arcsec \\
07/25.21--25.49	& 1.519 & 1.020 & 242 & H$_2$O($1_{10}-1_{01}$) & $0.30\pm0.06$ & $+0.15\pm0.15$ & 15\arcsec \\
07/31.59--31.87	& 1.529 & 1.064 & 207 & H$_2$O($1_{10}-1_{01}$) & $0.19\pm0.05$ & $-0.38\pm0.22$ & 20\arcsec \\
08/07.55--07.82	& 1.543 & 1.116 & 201 & H$_2$O($1_{10}-1_{01}$) & $0.32\pm0.05$ & $-0.05\pm0.13$ & 15\arcsec \\
\hline
\multicolumn{8}{c}{IRAM~30-m} \\
\hline
05/04.76--05.04 & 1.625 & 0.711 & 460 &    HCN(1--0)	& $0.031\pm0.005$ & $-0.18\pm0.10$ & 2.0\arcsec \\
05/05.78--06.04 & 1.622 & 0.712 & 418 &    HCN(3--2)	& $0.097\pm0.012$ & $-0.08\pm0.06$ & 2.3\arcsec \\
05/06.80--07.04 & 1.618 & 0.712 &  78 &    HCN(3--2)	& $0.213\pm0.034$ & $-0.22\pm0.08$ & 2.0\arcsec \\
		&	&	& 178 &    HCN(1--0)	& $0.045\pm0.008$ & $-0.27\pm0.11$ & 3.0\arcsec \\
05/07.79--08.04 & 1.615 & 0.713 & 150 &    HCN(3--2)	& $0.112\pm0.017$ & $-0.09\pm0.07$ & 2.0\arcsec \\
		&	&	& 200 &    HCN(1--0)	& $0.028\pm0.006$ & $-0.14\pm0.13$ & 2.0\arcsec \\
05/08.79--09.04 & 1.611 & 0.713 & 202 &    HCN(3--2)	& $0.212\pm0.019$ & $-0.12\pm0.08$ & 2.0\arcsec \\
05/05.78--07.04 & 1.621 & 0.712 & 287 & H$_2$S($1_{10}-1_{01}$)& $0.030\pm0.008$ & $-0.28\pm0.13$ & 2.1\arcsec \\
05/05.78--09.04 & 1.615 & 0.713 & 818 & CH$_3$OH~157~GHz & $0.055\pm0.008$ & $-0.13\pm0.08$ & 2.8\arcsec \\
05/04.76--08.04 & 1.619 & 0.712 & 508 & CO(2--1)	& $<0.025$ 	& 		& 2.7\arcsec \\
05/08.79--09.04 & 1.611 & 0.713 & 202 & CS(5--4)	& $<0.045$ 	& 		& 2.5\arcsec \\
05/04.76--05.04 & 1.625 & 0.711 & 230 & H$_2$CO($3_{12}-2_{11}$)& $<0.062$ 	& 		& 2.0\arcsec \\
\hline
07/02.81--03.88 & 1.506 & 0.889 & 221 &    HCN(1--0)	& $0.035\pm0.008$ & $-0.03\pm0.14$ & 2\arcsec \\
		&	&	& 221 &    HCN(3--2)	& $0.149\pm0.036$ & $-0.19\pm0.18$ & 3\arcsec \\
07/04.66--04.76 & 1.506 & 0.897 &  76 &    HCN(1--0)	& $0.023\pm0.016$ & $+0.04\pm0.27$ & 6\arcsec \\
		&	&	&  76 &    HCN(3--2)	& $0.147\pm0.041$ & $+0.18\pm0.23$ & 6\arcsec \\
07/04.76--04.92 & 1.506 & 0.897 & 125 &    HCN(1--0)	& $0.045\pm0.012$ & $+0.04\pm0.27$ & 3\arcsec \\
		&	&	& 125 &    HCN(3--2)	& $0.145\pm0.026$ & $-0.07\pm0.10$ & 4\arcsec \\
07/05.68--05.92 & 1.506 & 0.902 & 200 &    HCN(1--0)	& $0.037\pm0.010$ & $-0.38\pm0.23$ & 2\arcsec \\
		&	&	& 190 &    HCN(3--2)	& $0.147\pm0.037$ & $-0.40\pm0.20$ & 4\arcsec \\
07/06.80--10.91 & 1.507 & 0.919 & 585 &    HCN(1--0)	& $0.015\pm0.005$ & $-0.25\pm0.20$ & 5\arcsec \\
		&	&	& 585 &    HCN(3--2)	& $0.093\pm0.035$ & $+0.14\pm0.23$ & 3\arcsec \\
07/02.81--03.88 & 1.506 & 0.889 & 161 & CH$_3$OH($3_0-2_0$)A$^+$  & $0.038\pm0.009$ & $-0.40\pm0.15$ & 2\arcsec \\
		&	&  	&     & CH$_3$OH($3_{-1}-2_{-1}$)E& $0.014\pm0.009$ &  & \\
		&	&  	&     & CH$_3$OH($3_0-2_0$)E	  & $0.015\pm0.009$ &  & \\
07/04.76--04.92 & 1.506 & 0.897 & 125 & CH$_3$OH($3_0-2_0$)A$^+$  & $0.046\pm0.009$ & $-0.36\pm0.18$ & 3\arcsec \\
		&	&  	&     & CH$_3$OH($3_{-1}-2_{-1}$)E& $0.038\pm0.009$ & $-0.06\pm0.19$  & \\
		&	&  	&     & CH$_3$OH($3_0-2_0$)E	  & $0.019\pm0.010$ & $-0.52\pm0.58$  & \\
07/05.68--05.92 & 1.506 & 0.902 & 190 & CH$_3$OH~145~GHz & $0.068\pm0.018$ & $-0.09\pm0.15$ & 2.5\arcsec \\
07/06.80--09.89 & 1.506 & 0.917 & 320 & CH$_3$OH~145~GHz & $0.046\pm0.019$ & $-0.90\pm0.50$ & 4\arcsec \\
07/02.81--03.87 & 1.506 & 0.889 & 221 & CO(2--1)	& $<0.066$ 	& 		& 2\arcsec \\
07/04.76--04.92 & 1.506 & 0.897 & 125 & CO(2--1)	& $<0.051$ 	& 		& 3\arcsec \\
07/06.80--08.91 & 1.506 & 0.913 & 265 & CS(5--4)	& $<0.095$ 	& 		& 3\arcsec \\
07/05.68--10.91 & 1.506 & 0.914 & 305 & H$_2$CO($3_{12}-2_{11}$)& $<0.080$ 	& 		& 3\arcsec \\
\hline
\multicolumn{8}{c}{CSO~10.4-m} \\
\hline
07/04.22--04.25 & 1.506 & 0.895 &  27 & CH$_3$OH~305~GHz  & $<0.260$ &  & 3\arcsec \\
07/04.25--04.35 & 1.506 & 0.895 &  69 & CH$_3$OH~305~GHz  & $0.252\pm0.069$ & $+0.23\pm0.25$ & 3\arcsec \\
07/05.06--05.18 & 1.506 & 0.899 & 107 & CH$_3$OH~305~GHz  & $<0.167$ &  & 3\arcsec \\
07/05.21--05.36 & 1.506 & 0.900 & 123 & HCN(4--3)         & $<0.141$ &  & 3\arcsec  \\
\hline
\end{tabular}
\end{center}
CH$_3$OH~157~GHz: sum of the three ($3_0-3_{-1}$)E,($4_0-4_{-1}$)E and ($5_0-5_{-1}$)E lines of CH$_3$OH.\\
CH$_3$OH~145~GHz: sum of the three ($3_0-2_0$)A$^+$,($3_{-1}-2_{-1}$)E and ($3_0-2_0$)E lines of CH$_3$OH.\\
CH$_3$OH~305~GHz: sum of the ($2_1-2_0$)A$^{-+}$ line at 304.2~GHz and ($4_1-4_0$)A$^{-+}$ line at 307.2~GHz\\
}
\end{table}

\begin{table}
{\footnotesize
\begin{center}
\caption[]{Simulations of outburst: evolution of ``apparent'' production rates}\label{tabsimul}
\begin{tabular}{l|cc|cc|cc|cc|cc}
\hline
Model:     &\multicolumn{2}{c}{1} 	& \multicolumn{2}{c}{2}		& 
	\multicolumn{2}{c}{3}	& \multicolumn{2}{c}{4}		& \multicolumn{2}{c}{5}  \\
$v_{exp}=$ &\multicolumn{2}{c}{0.75~\kms}&\multicolumn{2}{c}{0.75~\kms} 	&
	\multicolumn{2}{c}{0.35~\kms}&\multicolumn{2}{c}{0.35~\kms}&\multicolumn{2}{c}{0.35~\kms} \\
$\Delta t$:&\multicolumn{2}{c}{1~h} 	& \multicolumn{2}{c}{15~h} 	& 
	\multicolumn{2}{c}{1~h}	& \multicolumn{2}{c}{4~h}	& \multicolumn{2}{c}{10~h} \\
\hline
Line	& H$_2$O & HCN(3--2) & H$_2$O & HCN(3--2) & H$_2$O & HCN(3--2) & H$_2$O & HCN(3--2) & H$_2$O & HCN(3--2) \\
$\Delta Q_a/Q$ &  0.32 &  2.21  &  0.15 &  0.29	  & 0.06 & 1.25        & 0.11 & 0.76        & 0.12 & 0.40 \\
\hline
$t_--t_0$= & 1.5~h & 0.5~h &  4~h & 0.7~h &  5~h & 0.9~h &  6~h & 1.4~h &  8~h & 1.5~h \\
$t_p-t_0$= &  3~h  & 1.5~h & 14~h &   4~h & 14~h &   2~h & 15~h & 3.5~h & 20~h &   5~h \\
$t_+-t_0$= & 14~h  & 2.5~h & 30~h &  17~h & 35~h &   4~h & 35~h &   7~h & 38~h & 12.5~h \\
\hline
\end{tabular}
\end{center}
Note: $t_p$, $t_-$ and $t_+$ are the times at which the ``apparent'' increase of 
outgassing reaches its maximum ($\Delta Q_a$), $\Delta Q_a/2$ when rising 
and $\Delta Q_a/2$ when decreasing, respectively.
$t_0$ is the beginning of the outburst, which will have decreased by a factor 2
after $\Delta t$.
}
\end{table}


\begin{table}
{\scriptsize
\begin{center}
\caption[]{Molecular ``apparent'' production rates of comet 9P/Tempel~1}\label{tabqprod}
\begin{tabular}{lccrrr}
\hline
  UT date           & $<r_{h}>$ & $<\Delta>$ & $Q_{\rm app}$  & $Q_{\rm H_2O,ref}^{1}$ \\[0cm]
[mm/dd.dd$\pm$d.dd] &   [AU]    &   [AU]     & 		  & [$10^{27}$s$^{-1}$]   \\
\hline
\multicolumn{3}{l}{OH production rates (Nan\c{c}ay)} & $[10^{27}$~\mols] &  \\
\hline
03/11.6$\pm7.5$		&  1.89 & 0.98  &  $<6.0$      &  5.0**  \\
03/28.8$\pm8.0$		&  1.78 & 0.82  &  $4.2\pm1.0$ &  5.7**  \\
04/14.3$\pm7.5$ 	&  1.71 & 0.75  &  $4.3\pm0.9$ &  6.2**  \\  
04/27.3$\pm4.5$		&  1.65 & 0.72  &  $4.9\pm1.3$ &  6.7**  \\
05/07.3$\pm4.5$		&  1.62 & 0.71  &  $9.5\pm1.4$ &  6.9**  \\
05/17.8$\pm5.0$		&  1.58 & 0.72  &  $6.2\pm1.9$ &  7.3**  \\
06/01.3$\pm7.5$		&  1.54 & 0.76  &  $6.0\pm1.5$ &  7.7**  \\
07/06.28$\pm4.5$	&  1.51	& 0.90	&   $<37.0$    &  8.0**  \\
\hline
\multicolumn{3}{l}{H$_2$O production rates (Odin)} & $[10^{27}$~\mols] & & \\
\hline
06/18.25$\pm0.18$	& 1.516 & 0.818 & $11.5\pm1.0$ &  7.9**  \\
06/23.73$\pm0.12$	& 1.511 & 0.842 & $12.4\pm1.3$ &  8.0**  \\
07/03.93$\pm0.28$	& 1.506 & 0.892 &  $9.1\pm0.8$ &  8.0**  \\
07/04.44$\pm0.18$   	& 1.506 & 0.895 &  $7.8\pm0.9$ &  8.0**  \\
07/04.84$\pm0.18$	& 1.506 & 0.897 &  $9.6\pm1.0$ &  8.0**  \\
07/05.24$\pm0.18$	& 1.506 & 0.899 & $10.8\pm1.0$ &  8.0**  \\
07/05.64$\pm0.19$	& 1.506 & 0.901 &  $8.7\pm1.1$ &  8.0**  \\
07/07.70$\pm0.14$	& 1.506 & 0.912 & $10.6\pm1.5$ &  8.0**  \\
07/09.34$\pm0.12$	& 1.507 & 0.921 &  $7.8\pm1.5$ &  8.0**  \\
07/10.71$\pm0.14$	& 1.507 & 0.929 &  $7.3\pm1.3$ &  8.0**  \\
07/12.07$\pm0.12$	& 1.508 & 0.937 &  $9.0\pm1.3$ &  8.0**  \\
07/16.29$\pm0.12$	& 1.510 & 0.962 &  $7.7\pm1.4$ &  8.0**  \\
07/25.35$\pm0.14$	& 1.519 & 1.020 &  $6.8\pm1.3$ &  8.0**  \\
07/31.73$\pm0.14$	& 1.529 & 1.064 &  $4.8\pm1.3$ &  7.8**  \\
08/07.68$\pm0.13$	& 1.543 & 1.116 &  $8.4\pm1.4$ &  7.6**  \\
\hline		 
\multicolumn{3}{l}{HCN production rates} & $[10^{24}$~\mols] & [$10^{27}$s$^{-1}$] & \\	
\hline
05/04.90$\pm0.14$ 	& 1.625 & 0.711 &  $6.8\pm1.1$ &  7.6*  \\
05/05.91$\pm0.13$ 	& 1.622 & 0.712 &  $4.7\pm0.6$ &  5.0*  \\
05/06.92$\pm0.12$ 	& 1.618 & 0.712 & $10.1\pm1.2$ &  9.5*  \\
05/07.92$\pm0.13$ 	& 1.615 & 0.713 &  $5.6\pm0.8$ &  4.5*  \\
05/08.92$\pm0.13$ 	& 1.611 & 0.713 & $10.1\pm1.4$ &  8.8*  \\
\hline
07/03.34$\pm0.53$ 	& 1.506 & 0.889 & $11.1\pm1.8$ &  8.0**  & \\
07/04.71$\pm0.05$ 	& 1.506 & 0.897 & $13.3\pm6.4$ &  impact & \\
07/04.84$\pm0.08$ 	& 1.506 & 0.897 & $13.1\pm2.1$ &  impact & \\
07/05.30$\pm0.03$	& 1.506 & 0.899 & $<84$        &  8.0**  & \\
07/05.80$\pm0.12$ 	& 1.506 & 0.902 & $12.3\pm2.3$ &  8.0**  & \\
07/08.85$\pm2.05$ 	& 1.507 & 0.919 &  $6.0\pm1.8$ &  8.0**  & \\
\hline
\multicolumn{3}{l}{CH$_3$OH production rates} & $[10^{25}$~\mols] & & \\
\hline
05/07.7$\pm1.6$ 	& 1.615 & 0.713 &  $18\pm3$ &   6.5*  \\
07/03.34$\pm0.53$ 	& 1.506 & 0.889 &  $25\pm6$ &   8.0** \\
07/04.23$\pm0.02$	& 1.506 & 0.894 &    $<88$  &   impact \\
07/04.30$\pm0.05$	& 1.506 & 0.895 &  $85\pm24$&   impact \\
07/04.84$\pm0.08$	& 1.506 & 0.897 &  $39\pm6$ &   impact \\
07/05.12$\pm0.06$	& 1.506 & 0.899 &    $<56$  &   8.0** \\
07/05.80$\pm0.12$	& 1.506 & 0.902 &  $20\pm7$ &   8.0** \\
07/06.80--09.89 	& 1.506 & 0.917 &  $18\pm6$ &   8.0** \\
\hline
\multicolumn{3}{l}{H$_2$S production rate} & $[10^{24}$~\mols]  & &\\
\hline
05/05.78--07.04		& 1.621 & 0.712 & $32\pm8$ & 6.5* \\
 \hline
\multicolumn{3}{l}{CS production rate upper limits} & $[10^{24}$~\mols]  & & \\
\hline
05/08.92$\pm0.13$ 	& 1.611 & 0.713 & $<11.1$ & 8.8*  \\
07/06.80--8.91		& 1.506 & 0.913 & $<19.7$ & 8.0** \\ 
\hline
\multicolumn{3}{l}{CO production rate upper limits} & $[10^{26}$~\mols]  & & \\
\hline
05/04.76--8.04		& 1.619 & 0.712 & $<7.4$  & 7.2*  \\
07/03.34$\pm0.53$ 	& 1.506 & 0.889 & $<25.5$ & 8.0** \\
07/04.84$\pm0.08$	& 1.506 & 0.897 & $<20.0$ & impact \\
\hline
\multicolumn{3}{l}{H$_2$CO production rates upper limits} & $[10^{24}$~\mols]  & & \\
\hline
05/04.90$\pm0.14$ 	& 1.625 & 0.711 & $<115$ & 7.6*  \\ 
07/05.68--10.91		& 1.506 & 0.914 & $<184$ & 8.0**  \\ 
\hline
\end{tabular}
\end{center}
$^1$ $Q_{\rm H_2O,ref}$ is from Eq.~(\ref{fqrefiram}) (*), or from 
($8\times10^{27}\times(1.506/r_h)^2$, cf. Sect. 4.4) 
for non IRAM May data (**) -- which is also Eq.~(\ref{fqrefiram}) 
removing the sine term (cf. Sect. 3.2). 
In case of several days averages, the given $Q_{\rm H_2O,ref}$ is also the 
weighted average of each day value.\\  
}
\end{table}

\begin{table}
\begin{center}
\caption[]{Molecular abundances of comet 9P/Tempel~1}\label{tababund}
\begin{tabular}{lcc}
\hline
Molecule	& $Q_{\rm molec.}/Q_{\rm H_2O,ref}$ & $Q_{\rm molec.}/Q_{\rm H_2O,ref}$ \\
		& May 2005			   & 2--3 or 5--10 July 2005	\\
\hline
HCN		& $0.11\pm0.01$ \%	& $0.12\pm0.03$ \%	\\
CH$_3$OH	& $2.8\pm0.9$ \%	& $2.7\pm0.4$ \%	\\
H$_2$S		& $0.5\pm0.1$ \%	&			\\
CS$^1$		& $<0.13$ \%		& $<0.25$ \%		\\
CO		& $<10$ \%		& $<32$ \%		\\
H$_2$CO$^2$	& $<1.5$ \%		& $<2.3$ \%		\\
\hline
\end{tabular}
\end{center}
$^1$ CS is assumed to come from CS$_2$ (600~km parent scalelength used here).\\
$^2$ H$_2$CO is assumed to come from a distributed source with a lifetime 
equals to 1.75 times the H$_2$CO lifetime (Biver et al. 1999).\\
\end{table}

\begin{table}\hspace{-2cm}
{\footnotesize
\begin{center}
\caption[]{Rotationally phased H$_2$O production rates (July 2005)}\label{tabqh2opha}
\begin{tabular}{l|lrr|lccrrr}
\hline
Phase$^1$ & \multicolumn{3}{c}{Data around impact}   & \multicolumn{5}{c}{``Normal'' activity 3 to 34 days after impact} \\[0cm]
          & UT date & $\int T_bdv$ & $Q_{\rm H_2O}$     & Date  & $<r_{h}>$ & $<\Delta>$ & $\int T_bdv$    & dv & ${Q_{\rm H_2O}}^2$ \\[0cm]
         & [dd.dd] & [K~km~s$^{-1}$] & [$10^{27}$s$^{-1}$] & [dd.dd] &  [AU] &  [AU]  & [K~km~s$^{-1}$] & [\kms] & [$10^{27}$s$^{-1}$]   \\
\hline
$-0.18$ & 03.93 & $0.44\pm0.04$ &  $9.1\pm0.6$ & 09.33+10.71  & 1.507 & 0.925 & $0.34\pm0.04$ & $-0.08\pm0.10$ & $7.3\pm0.9$ \\
$+0.12$ & 04.44 & $0.38\pm0.04$ &  $7.8\pm0.9$ & 07.72+31.73  & 1.515 & 0.979 & $0.32\pm0.03$ & $-0.19\pm0.09$ & $7.4\pm0.7$ \\
$+0.35$ & 04.84 & $0.46\pm0.05$ &  $9.6\pm1.0$ & 25.37+38.68  & 1.531 & 1.068 & $0.30\pm0.04$ & $-0.03\pm0.12$ & $7.6\pm1.0$ \\
$+0.58$ & 05.24 & $0.52\pm0.05$ & $10.8\pm1.0$ & 12.07        & 1.508 & 0.937 & $0.42\pm0.06$ & $-0.21\pm0.16$ & $9.0\pm1.3$ \\
$+0.82$ & 05.65 & $0.41\pm0.05$ &  $8.7\pm1.1$ & 09.33+10.71  & 1.507 & 0.925 & $0.34\pm0.04$ & $-0.08\pm0.10$ & $7.3\pm0.9$ \\
\hline
\end{tabular}
\end{center}
$^1$ Rotation phase with reference time $t_0$= 4.244 July 2005 (impact date) 
and a 1.7~days period.\\
$^2$ Corrected for a priori $r_h^{-2}$ variation since impact day ($r_h=1.506$ AU).
}
\end{table}

\begin{table}
\begin{center}
\caption[]{Impact residual H$_2$O production rate}\label{tabqh2oimp}
\begin{tabular}{lc|cc}
\hline
UT date    & $\Delta~Q_{\rm app.H_2O}$ & ``Spike'' area & $(\Delta~Q_{\rm H_2O})_{\rm app}$\\[0cm]
           &  model (1)            & $\int T_bdv$ & model (2)  \\[0cm]
[mm/dd.dd] &  [$10^{27}$s$^{-1}$]  & [K~km~s$^{-1}$] & [$10^{27}$s$^{-1}$]\\
\hline
07/04.44 & $0.4\pm1.1$ & $0.02\pm0.05$ & $0.2\pm0.5$ \\
07/04.84 & $2.0\pm1.5$ & $0.10\pm0.06$ & $0.9\pm0.5$ \\
07/05.24 & $1.8\pm1.6$ & $0.09\pm0.07$ & $0.8\pm0.6$ \\
07/05.65 & $1.4\pm1.4$ & $0.07\pm0.06$ & $0.7\pm0.5$ \\
\multicolumn{4}{c}{Average} \\
07/05.04 & $1.4\pm0.7$ & $0.08\pm0.02$ & $0.7\pm0.2$ \\
\hline
\multicolumn{4}{c}{Integrated mass of water $[tons]$: 07/04.24--05.83} \\
\hline
07/05.04 & $5700\pm2800$ &  & $5300\pm1800^1$  \\
\hline
\end{tabular}
\end{center}
Notes: \\
Model (1): $Q_{\rm H_2O}$(Table~\ref{tabqprod}) minus $Q_{\rm H_2O}$ (Table~\ref{tabqh2opha} for same phase).\\
Model (2): spike area based on the difference of line 
integrated intensities from Table~\ref{tabqh2opha} after corrections for all 
geometrical variation effects ($r_h$, $\Delta$, $v_{exp}$, photo-dissociation rates...) 
on line intensities. \\ 
$^1$ Integrated $\Delta~Q_{\rm H_2O}~dt$ has been multiplied by 2 
since only 30--70\% of the molecules are seen in the hypothesis of a low speed,
1--10~h duration jet (cf. Section 3.2).
\end{table}

\clearpage

\vspace{-1cm}

\vspace{-0cm}

\vspace{-1cm}

\end{document}